\documentclass[twocolumn,showpacs,preprintnumbers,amsmath,amssymb,superscriptaddress]{revtex4-1}

\usepackage{graphicx}
\usepackage{dcolumn}
\usepackage{bm}
\usepackage{tabularx}
\usepackage{makecell}
\usepackage{braket}
\usepackage{lipsum}
\usepackage{ulem}
\usepackage{amsmath,amssymb} 
\newcolumntype{M}{>{\centering\arraybackslash}m{1.85cm}}
\usepackage[export]{adjustbox}
\usepackage{longtable}
\usepackage{float}
\usepackage{xcolor}
\usepackage{color}   
\usepackage{hyperref}
\hypersetup{
	colorlinks=true, 
	linktoc=all,     
	linkcolor=blue,  
}

\makeatletter
\newcommand{\colorcaption}[2][]{%
	\begingroup%
	\renewcommand{\@caption@fignum@sep}{ (Color online). }%
	\caption[#1]{#2}%
	\endgroup%
}
\makeatother
\newcommand{\orcid}[1]{\href{https://orcid.org/#1}{\hskip2pt\includegraphics[width=9pt]{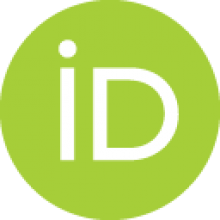}}}
\begin{document}
	
	
	
	\title{Evaluation of $2\nu\beta\beta$ decay nuclear matrix elements using Lanczos techniques preserving Gamow-Teller strengths from both parent and daughter nuclei}
	
	\author{Deepak Patel\orcid{0000-0002-7669-1907}}
	\email{patel@cns.s.u-tokyo.ac.jp}
	\address{Center for Nuclear Study, University of Tokyo, 7-3-1 Hongo, Bunkyo-ku, Tokyo 113-0033, \\Japan}
	\author{Anil Kumar\orcid{0000-0003-3299-0435}}
	\email{anil@nucl.ph.tsukuba.ac.jp}
	\address{Center for Computational Sciences, University of Tsukuba, Tennodai, Tsukuba, Ibaraki 305-8577, Japan}
	\author{Noritaka Shimizu\orcid{0000-0003-1638-1066}}
	\email{shimizu@nucl.ph.tsukuba.ac.jp}
	\address{Center for Computational Sciences, University of Tsukuba, Tennodai, Tsukuba, Ibaraki 305-8577, Japan}
	\address{Center for Nuclear Study, University of Tokyo, 7-3-1 Hongo, Bunkyo-ku, Tokyo 113-0033, \\Japan}
	\author{Yutaka Utsuno\orcid{0000-0001-5404-7320}}
	\email{utsuno.yutaka@jaea.go.jp}
	\address{Advanced Science Research Center, Japan Atomic Energy Agency, Tokai, Ibaraki 319-1195, Japan}
	\address{Center for Nuclear Study, University of Tokyo, 7-3-1 Hongo, Bunkyo-ku, Tokyo 113-0033, \\Japan}

	\date{\hfill \today}
	
	
	\begin{abstract}				
		
		We present two efficient shell-model approaches for calculating nuclear matrix elements (NMEs) for two-neutrino double-beta ($2\nu\beta\beta$) decay and two-neutrino double-electron capture ($2\nu$ECEC), based on the Lanczos strength-function and block-Lanczos techniques. The dual-sided Lanczos strength-function (DLSF) and block-Lanczos strength-function (BLSF) methods avoid the explicit calculation of a complete set of exact intermediate $1^+$ eigenstates and yield fully converged NMEs with a relatively small number of iterations starting from Gamow--Teller (GT) doorway states. Both methods simultaneously preserve the total GT strengths associated with the initial and final nuclei. We benchmark the methods for $^{48}$Ca in the $fp$ shell and apply them to candidates in different mass regions, including $^{58}$Ni, $^{76}$Ge, and $^{130}$Te. For $^{48}$Ca, a calculation allowing cross-shell excitations of up to $2\hbar\omega$ in the $sdpf$ model space yields a $2\nu\beta\beta$ NME closer to the experimental estimate than the result obtained in the $fp$ shell. For $^{58}$Ni, we present, to our knowledge, the first shell-model calculation of the $2\nu\beta^+$EC/$2\nu$ECEC NME. Comparison with the current experimental upper limit on the NME indicates the need for a more sensitive measurement of the $2\nu\beta^+$EC half-life. As an independent validation, the energy-denominator-free sums obtained with the proposed methods are compared with the corresponding double-GT matrix elements calculated directly between the initial and final ground states. Their agreement demonstrates that the methods correctly recover the complete energy-denominator-free intermediate-state sums.
		
	\end{abstract}
	
	\maketitle
	
	
	\section{\label{sec:level1}Introduction}
	
	The nature of the neutrino and its possible role in the matter-antimatter asymmetry of the Universe remain among the fundamental open questions in nuclear and particle physics \cite{Cadenas, Agostini}. If neutrinos are Majorana particles, neutrinoless double-beta ($0\nu\beta\beta$) decay can occur, which violates lepton-number conservation and provides a unique probe of the absolute neutrino-mass scale \cite{Iwata}. In contrast, having been experimentally observed, the allowed two-neutrino ($2\nu$) mode provides a vital benchmark for nuclear-structure models and an important background for $0\nu\beta\beta$ searches \cite{Ackerman, Adams}. In both modes, reliable nuclear matrix elements (NMEs), which connect the initial and final nuclear states through virtual states of the intermediate nucleus, are indispensable for interpreting experimental data \cite{Jokiniemi, Engel}. The nuclear shell model provides a microscopic and well-established framework for calculating such NMEs \cite{Horoi_136Xe, Kostensalo_48Ca}, but its applicability to medium- and heavy-mass nuclei is severely constrained by the rapid growth of the Hamiltonian matrix dimension in the adopted model space. For $2\nu\beta\beta$ decay, a direct evaluation requires a sum over the $1^+$ states of the intermediate nucleus, making a complete treatment increasingly computationally prohibitive as the model space grows. Consequently, conventional shell-model calculations have often been restricted to a limited number of intermediate states or to truncated model spaces \cite{Kostensalo_76Ge, Patel_76Ge}, potentially leaving a substantial fraction of the Gamow-Teller (GT) strength unaccounted for and thereby affecting the convergence and reliability of the calculated $2\nu\beta\beta$ NMEs. This limitation becomes particularly important because GT strength can extend to relatively high excitation energies. With increasingly precise $\beta\beta$ decay experiments \cite{Adams, Agostini_76Ge, Yan124Xe}, computationally efficient methods capable of incorporating the GT strength over a broad energy range are therefore highly desirable.

	In this work, we develop two new approaches within the shell-model framework, based on the Lanczos strength-function (LSF) \cite{Caurier_Review} and block-Lanczos procedures \cite{KShell}, for an efficient and accurate evaluation of $2\nu\beta\beta$ NMEs. Both approaches avoid the explicit calculation of the complete set of intermediate $1^+$ eigenstates while retaining essentially the full GT strength accessible within the adopted model space.  These approaches significantly reduce the computational cost of conventional state-by-state summation without introducing severe model-space truncations or approximations to the intermediate states. The present framework treats the relevant GT transitions entirely within the shell-model many-body space. In contrast to quasiparticle random-phase approximation (QRPA) based approaches, which describe intermediate states primarily in terms of one-particle-one-hole ($1p$-$1h$) excitations \cite{Civitarese_QRPA, Muto}, it incorporates correlations beyond the $1p$-$1h$ level within the adopted model space and thus provides a complementary perspective on the NMEs.
	
	The LSF method has previously been applied to the $2\nu\beta\beta$ decay of $^{48}$Ca in Ref. \cite{Caurier_LSF} and subsequently in Ref. \cite{Horoi_LSF}. In the pioneering LSF treatment \cite{Caurier_LSF}, however, the generated $1^+$ strength distribution was associated with a single GT doorway, i.e., constructed from either the initial or final nucleus. The approaches developed in the present work provide a more symmetric and robust treatment by incorporating the GT strength associated with both the initial and final nuclei in a more consistent way. In particular, the resulting $1^+$ states preserve the GT strengths from both the initial and final nuclei, which were not simultaneously enforced in the previous work \cite{Caurier_LSF}. This feature enables a consistent treatment of the intermediate-state contributions while retaining the computational efficiency of our approaches, providing an accurate and practical framework for calculating $2\nu\beta\beta$ NMEs in large shell-model spaces.

	The present study aims to benchmark both of our approaches for $2\nu\beta\beta$ decay in candidate nuclei across different mass regions and to assess their computational efficiency. Particular attention is devoted to obtaining a precise $2\nu\beta\beta$ NME for $^{48}$Ca in the $sdpf$ shell and a $2\nu$ double electron capture (2$\nu$ECEC) NME for $^{58}$Ni in the $fp$ shell. Furthermore, the energy-denominator-free sum entering the $2\nu\beta\beta/2\nu$ECEC NME is compared with the corresponding exact double-Gamow-Teller (DGT) transition matrix element, which can be obtained independently through a closure relation and therefore provides a stringent benchmark for the present approaches.
	
	This paper is organized as follows: In Sec. \ref{sec:level2}, we briefly present an overview of the formalism employed in our calculations. In Sec. \ref{sec:level3}, both of our methods are benchmarked in the $fp$ shell for $^{48}$Ca, and in Sec. \ref{sec:level4}, they are applied to several $2\nu\beta\beta/2\nu$ECEC candidates across different mass regions. In Sec. \ref{sec:level5}, the energy-denominator-free sums of the NMEs from our approaches are compared with the exact DGT matrix elements. Finally, the results are summarized and conclusions are drawn in Sec. \ref{sec:level6}.

	\section{\label{sec:level2}Theoretical Framework}
	
	The $2\nu\beta\beta/2\nu$ECEC decay NME between the $0^+$ ground states of the initial and final nuclei can be expressed as \cite{Stefanik, Caurier}
	\begin{equation}
		M_{2\nu}= \sum_{n}\frac{\langle 0_{\rm g.s.}^{+(f)}||\sum_{a}\boldsymbol{\sigma}_a\tau_a^\mp||1_n^+\rangle \langle 1_n^+||\sum_{b}\boldsymbol{\sigma}_b\tau_b^\mp||0_{\rm g.s.}^{+(i)}\rangle}{[E(1^+_n)-(E_i+E_f)/2]/m_e}.
		\label{eq2}
	\end{equation}
	
	Here, $E(1_n^+)$ denotes the energy of the $n$-th $1^+$ state in the intermediate nucleus. The states $0_{\rm g.s.}^{+(i)}$ and $0_{\rm g.s.}^{+(f)}$ represent the ground states of the initial and final nuclei, respectively, with corresponding energies $E_i$ and $E_f$. The electron rest mass is denoted by $m_e$. The Pauli spin matrices are denoted by $\boldsymbol{\sigma}_a$ and $\boldsymbol{\sigma}_b$, while $\tau_a^\mp$ and $\tau_b^\mp$ represent the isospin operators for neutron-to-proton ($\tau^-$) and proton-to-neutron ($\tau^+$) transitions. Thus the reduced GT matrix elements are given by $\langle 0_{\rm g.s.}^{+(f)} || \sum_a \boldsymbol{\sigma}_a \tau_a^\mp || 1_n^+ \rangle$ and $\langle 1_n^+ || \sum_b \boldsymbol{\sigma}_b \tau_b^\mp || 0_{\rm g.s.}^{+(i)} \rangle$. In our calculations, a common energy shift is applied to all intermediate $1^+$ states to reproduce the experimental energy of the lowest $1^+$ state \cite{NNDC_ENSDF}, except for $^{130}$Te, for which the shell-model-predicted energies are used without a shift because the experimental energy of the lowest $1^+$ state in $^{130}$I is not identified.
	
	The half-life for the $2\nu\beta\beta$ decay is expressed as
	\begin{equation} \label{eq1}
		T_{1/2}^{2\nu}=\frac{1}{G^{2\nu}(g_{\rm A})^4|M_{2\nu}|^2}.
	\end{equation}
	
	Here, $G^{2\nu}$ represents the phase-space factor \cite{Neacsu}, while $g_{\rm A}$ corresponds to the axial-vector coupling strength \cite{Suhonen_ga_eff}.

	As mentioned in Sec. \ref{sec:level1}, both of our approaches preserve the GT strengths for the initial and final nuclei; hence, we denote the first method as the dual-sided Lanczos strength-function (DLSF) method. Although the second, based on block-Lanczos (BL), also preserve the GT strengths for both nuclei, we denote it simply as the block-Lanczos strength-function (BLSF) method. The method first applied by Caurier $et$ $al.$ \cite{Caurier_LSF} to $2\nu\beta\beta$ decay is denoted as the single-sided Lanczos strength-function (SLSF) method. These notations are used throughout the paper. In DLSF and BLSF approaches for calculating the $2\nu\beta\beta/2\nu$ECEC NME, we first define the starting vectors $|L^{\pm}_{\rm start}\rangle$ by applying the GT operator to the initial and final $0^+$ ground states, respectively, as follows
	\begin{equation} \label{eq-3}
		\begin{aligned}
			\boldsymbol{\sigma} \tau^- |0_{\rm g.s.}^{+(i)} \rangle  \equiv s_- |L_{\rm start}^- \rangle \\
			\boldsymbol{\sigma} \tau^+ |0_{\rm g.s.}^{+(f)} \rangle  \equiv s_+ |L_{\rm start}^+ \rangle.
		\end{aligned}
	\end{equation}
	
	Here, $s_{\mp}$ are defined as the square roots of the corresponding total GT strengths
	\begin{equation}
		\label{eq-4-spm}
		\begin{aligned}
			s_- &= \sqrt{\sum_n \bigl|
				\langle 1_n^+ \bigl\| \sum_a \boldsymbol{\sigma}_a\tau_a^-
				\| 0_{\rm g.s.}^{+(i)} \rangle \bigr|^2}\\
			s_+ &= \sqrt{\sum_n \bigl|
				\langle 1_n^+ \bigl\| \sum_a \boldsymbol{\sigma}_a\tau_a^+
				\| 0_{\rm g.s.}^{+(f)} \rangle \bigr|^2}.
		\end{aligned}
	\end{equation}
	
	Here, the summation runs over $1^+$ states in the
	intermediate nucleus. After constructing the two starting vectors above, the following two formulations are adopted for the respective methods.

	\subsection{\label{sec:LSF_based}DLSF method}
	
	In this approach, we perform a limited number of Lanczos iterations for both initial vectors in Eq.~(\ref{eq-3}), generating two sets of an equal number of $1^+$ states for the intermediate nucleus. Although these states are not exact eigenstates, they define an approximate strength function that converges toward the full GT strength distribution while preserving the GT strengths associated with the initial and final nuclei. The $1^+$ states generated from the $\boldsymbol{\sigma}\tau^-$ and $\boldsymbol{\sigma}\tau^+$ channels are treated separately, then combined and orthogonalized using L\"owdin orthogonalization \cite{Torun}. In the L\"owdin orthogonalization, we first combine the two sets, each containing $n$ vectors, into a single set of $2n$ vectors, denoted by $|\phi_i\rangle$ ($i=1,2,\ldots,2n$). We then construct the $2n\times2n$ overlap matrix $S$, with elements $S_{ij}=\langle\phi_i|\phi_j\rangle$, where $i,j=1,2,\ldots,2n$. The overlap matrix $S$ is real and symmetric in the present calculations, and is diagonalized as $S = U\Lambda U^T$, where $U$ and $\Lambda$ are the eigenvector and diagonal eigenvalue matrices, respectively, with $\Lambda=\mathrm{diag}(\lambda_1,\lambda_2,\ldots,\lambda_{2n})$. We adopt the canonical L\"owdin orthogonalization, which provides an orthonormal basis and improves numerical stability by removing states associated with very small overlap eigenvalues. Eigenvalues ($\lambda$'s) smaller than $10^{-4}$ are therefore discarded, and the remaining orthogonalized vectors are written as
	\begin{equation}
		|\psi_k\rangle=\sum^{2n}_{j=1} \frac{U_{jk}}{\sqrt{\lambda_k}}|\phi_j\rangle,
	\end{equation}
	where $k=1,2,\ldots,n'$ and $n'\leq 2n$. Since the L\"owdin-orthogonalized states are no longer associated with the original Lanczos energies, including that of the lowest-lying state, the Hamiltonian can be rediagonalized in the subspace spanned by the L\"owdin states to construct the Ritz vectors. The corresponding Hamiltonian matrix elements are given by
	\begin{equation}
		H_{ij}=\langle \psi_i|H|\psi_j\rangle = \langle \psi_i|\psi'_j\rangle,
	\end{equation}
	where $i,j=1,2,\ldots,n'$ and $|\psi'_j\rangle=H|\psi_j\rangle$ is the state obtained by applying the Hamiltonian operator $H$ to the L\"owdin-orthogonalized state $|\psi_j\rangle$. Diagonalizing this Hamiltonian matrix, we obtain the eigenvector matrix $U'$, whose elements $U'_{ji}$ are the components of the $i$-th normalized eigenvector in the L\"owdin-orthogonalized basis. The Ritz vectors are then constructed as \cite{Yoshida}
	\begin{equation}
		|\psi_i^{\rm R1}\rangle = \sum_{j=1}^{n'} U'_{ji}|\psi_j\rangle,
	\end{equation}
	where $i=1,2,\ldots,n'$. These Ritz vectors $|\psi^{\rm R1}_i\rangle$ are then used as the intermediate $1^+$ states in the $2\nu\beta\beta/2\nu$ECEC NME calculation.

	\subsection{\label{sec:BL_based}BLSF method}
	
	In the diagonalization procedure for the shell-model Hamiltonian, when a large number of states are to be calculated, the BL method is an efficient approach to mitigate the computational bottleneck of on-the-fly matrix element generation by substantially decreasing the number of iterations, thereby effectively shortening the total elapsed time \cite{KShell}. In this approach, we first orthonormalize the initial vectors in Eq.~(\ref{eq-3}) using Gram-Schmidt orthogonalization before the BL iteration. The resulting orthonormalized vectors $|\tilde{\phi}_1\rangle$ and $|\tilde{\phi}_2\rangle$ form the initial block, $V_1=(|\tilde{\phi}_1\rangle,|\tilde{\phi}_2\rangle)$, with block size $q=n_{\rm block}=2$, satisfying $V_1^TV_1=\boldsymbol{I}$. The BL procedure then generates the block Krylov subspace \cite{KShell}
	\begin{equation}
		\mathcal{K}_m(H,V_1) = {\rm span}\{V_1, HV_1, H^2V_1, \ldots, H^{m-1}V_1\},
	\end{equation}
	where $H$ is the Hamiltonian in the subspace of the intermediate $1^+$ states. At each block iteration, the Hamiltonian acts simultaneously on the two vectors as $W_k=HV_k$, and the corresponding $2\times2$ block of the projected Hamiltonian is obtained as $\alpha_k=V^T_kHV_k$. The new block is then orthogonalized against the previously generated block vectors and factorized by a QR decomposition,
	\begin{equation}
		W_k-\sum^k_{j=1}V_j(V_j^TW_k)=V_{k+1}\beta_k,
	\end{equation}
	where $\beta_k$ is a $2\times2$ matrix.

	The Hamiltonian projected onto the block Krylov space therefore takes the symmetric block-tridiagonal form \cite{KShell}
	\begin{equation}
		T_m =
		\begin{pmatrix}
			\alpha_1 & \beta_1^T &  &  & 0 \\
			\beta_1 & \alpha_2 & \beta_2^T & \ddots &  \\
			& \beta_2 & \alpha_3 & \ddots &  \\
			& \ddots & \ddots & \ddots & \beta_{m-1}^T \\
			0 & &  & \beta_{m-1} & \alpha_m
		\end{pmatrix},
	\end{equation}
	whose dimension is $2m\times 2m$. Diagonalization of this relatively small matrix, $T_mU^i=\epsilon_iU^i$, provides the Ritz values $\epsilon_i$ and the corresponding Ritz vectors in the subspace of the intermediate $1^+$ states,
	\begin{equation}
		|\psi^{\rm R2}_i\rangle=\sum^{m}_{k=1} V_kU^i_k,
	\end{equation}
	where $U^i_k$ denotes the two-component part of the eigenvector $U^i$ belonging to the $k$-th block. Thus, after $m$ BL iterations, the calculation provides up to $2m$ Ritz vectors for intermediate $1^+$ states. Similar to the first approach, these $1^+$ states are also not exact eigenstates but progressively exhaust the full GT strength distribution as the number of iterations increases. Both methods are implemented in the KSHELL code \cite{KShell}, which is used for all calculations.

	\section{\label{sec:level3}Benchmark for $^{48}\mathrm{Ca}$ in $fp$ shell and computational efficiency}

	\begin{figure*}
		\includegraphics[width=45.0mm]{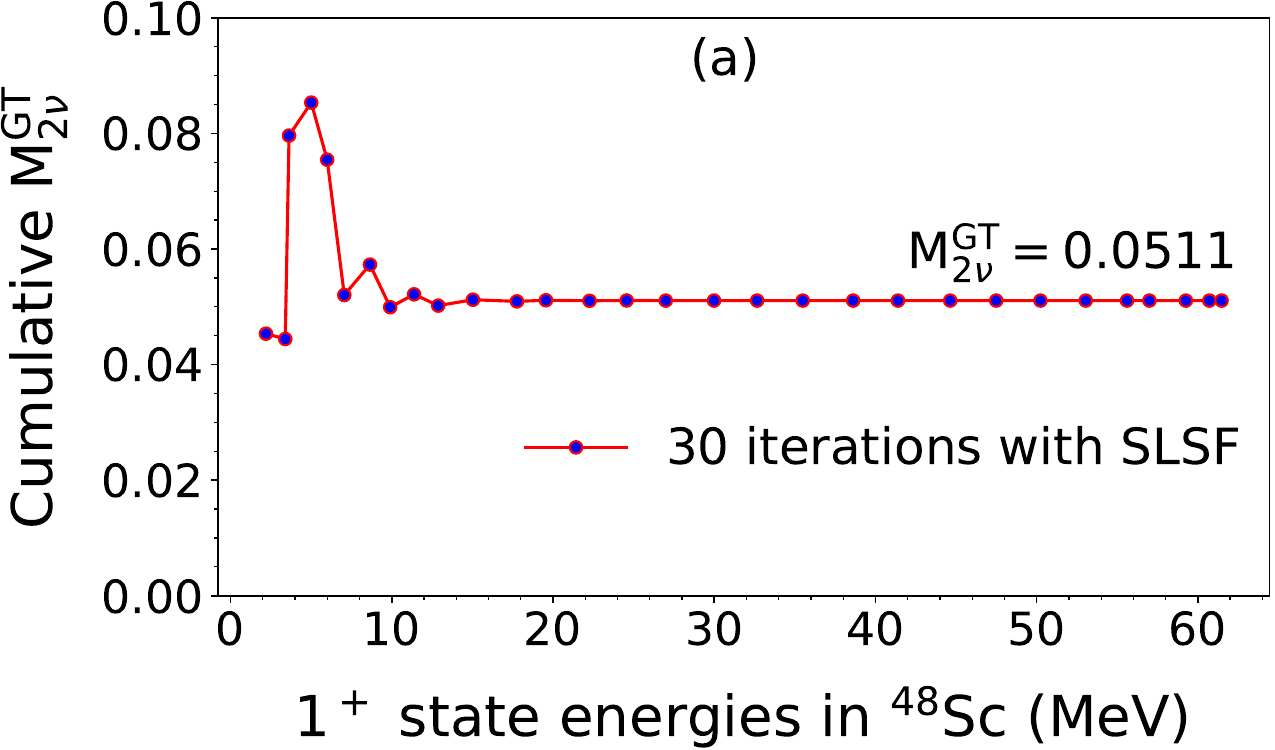}
		\includegraphics[width=41.8mm]{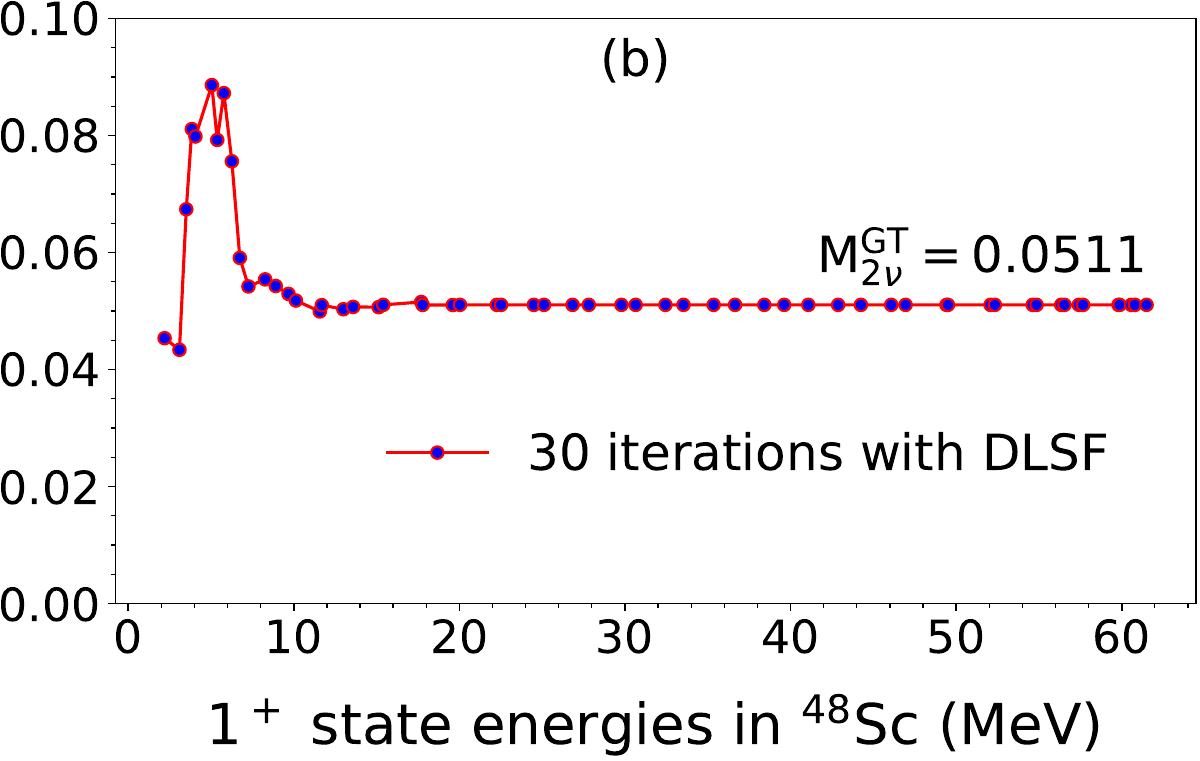}
		\includegraphics[width=41.8mm]{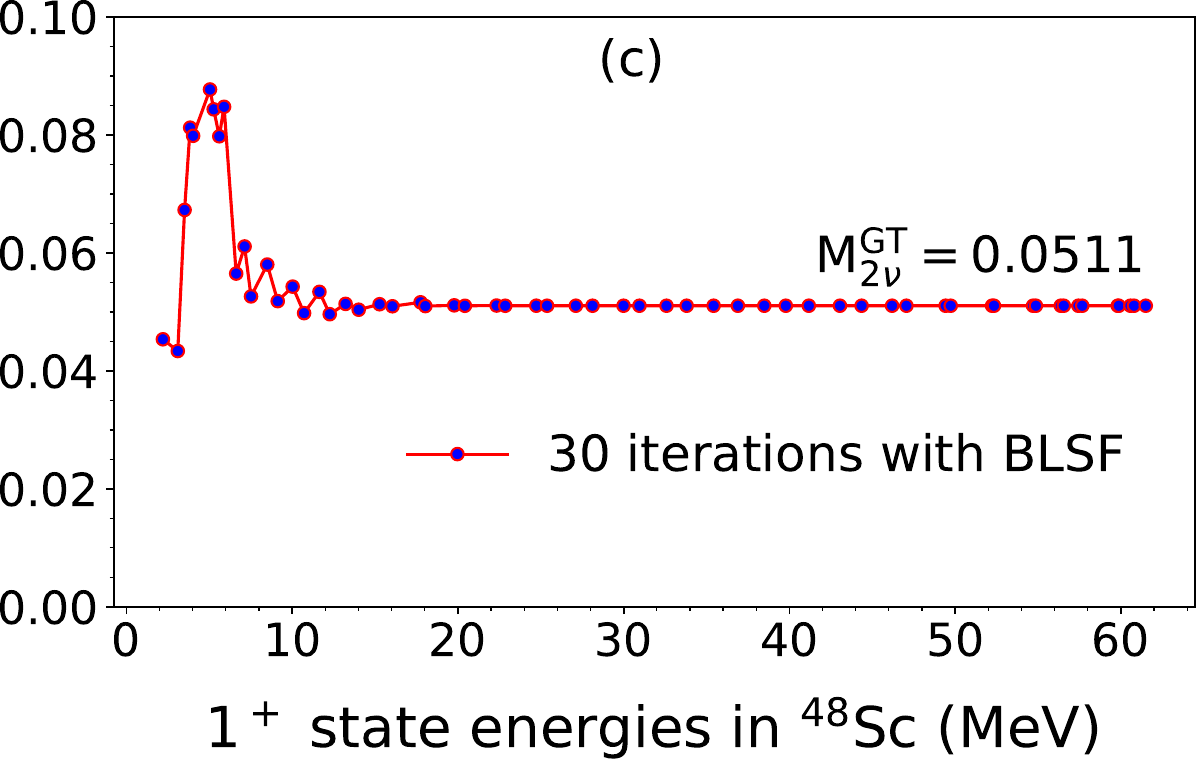}
		\includegraphics[width=41.8mm]{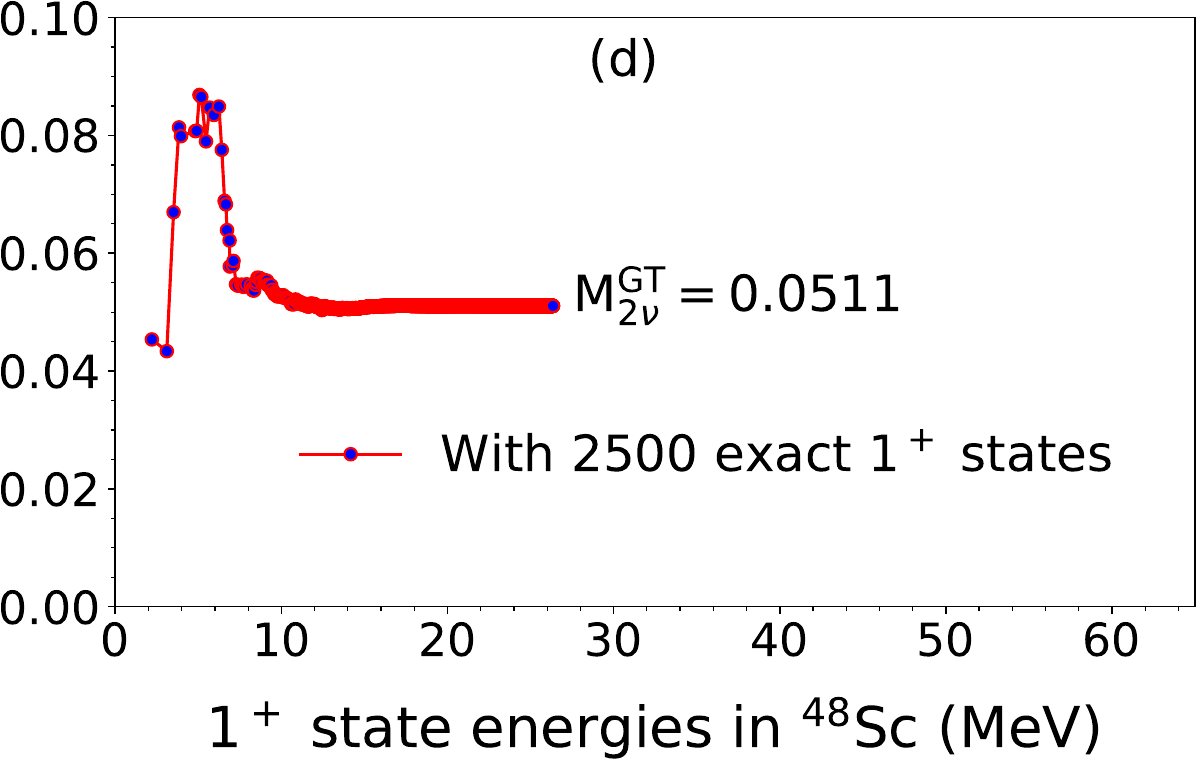}
		\caption{Comparison of the $2\nu\beta\beta$ NME for $^{48}$Ca with the GXPF1A interaction obtained using the (a) SLSF, (b) DLSF, (c) BLSF, and (d) exact-diagonalization methods.}
		\label{NME_48Ca_gxpf1a}
	\end{figure*}

	\begin{figure*}
		\includegraphics[width=50.0mm]{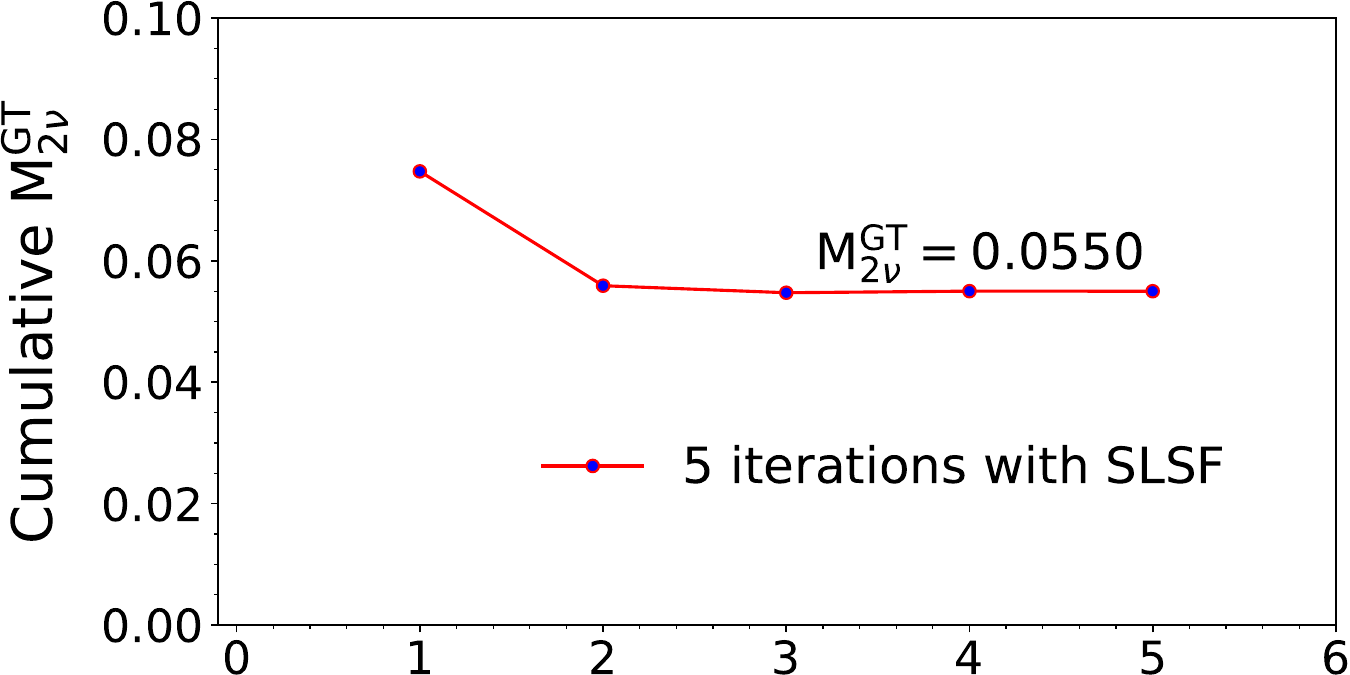}
		\includegraphics[width=46.8mm]{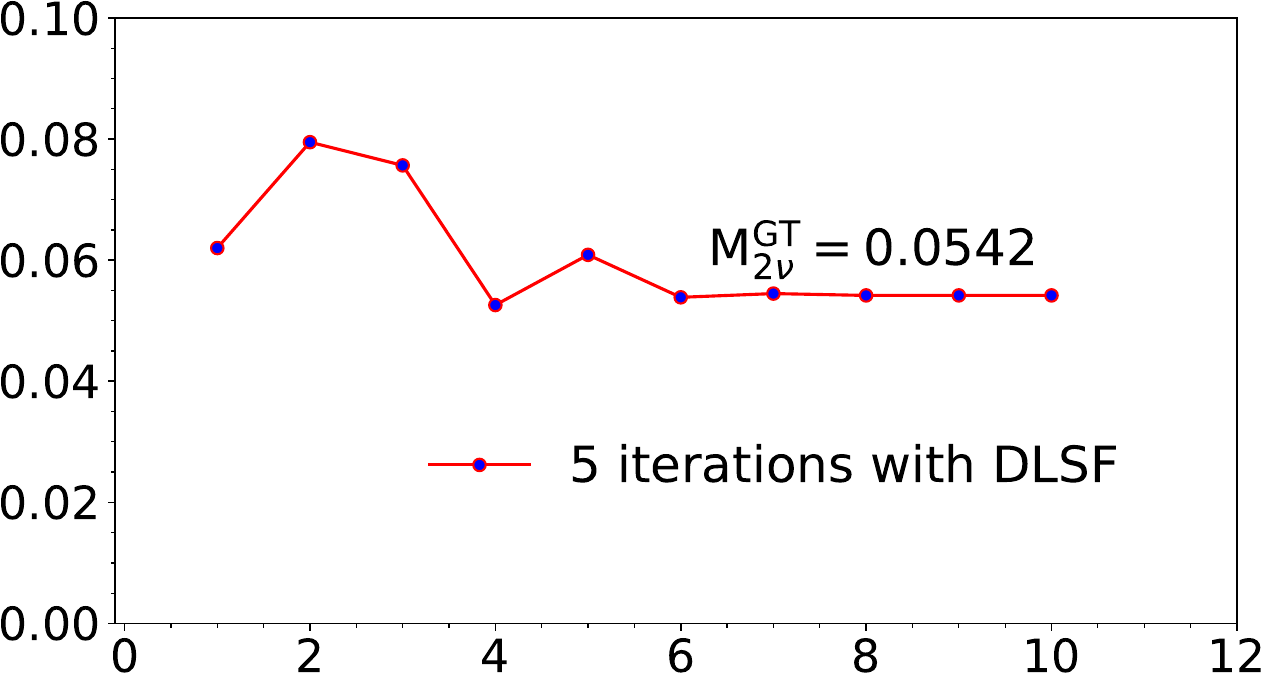}
		\includegraphics[width=46.8mm]{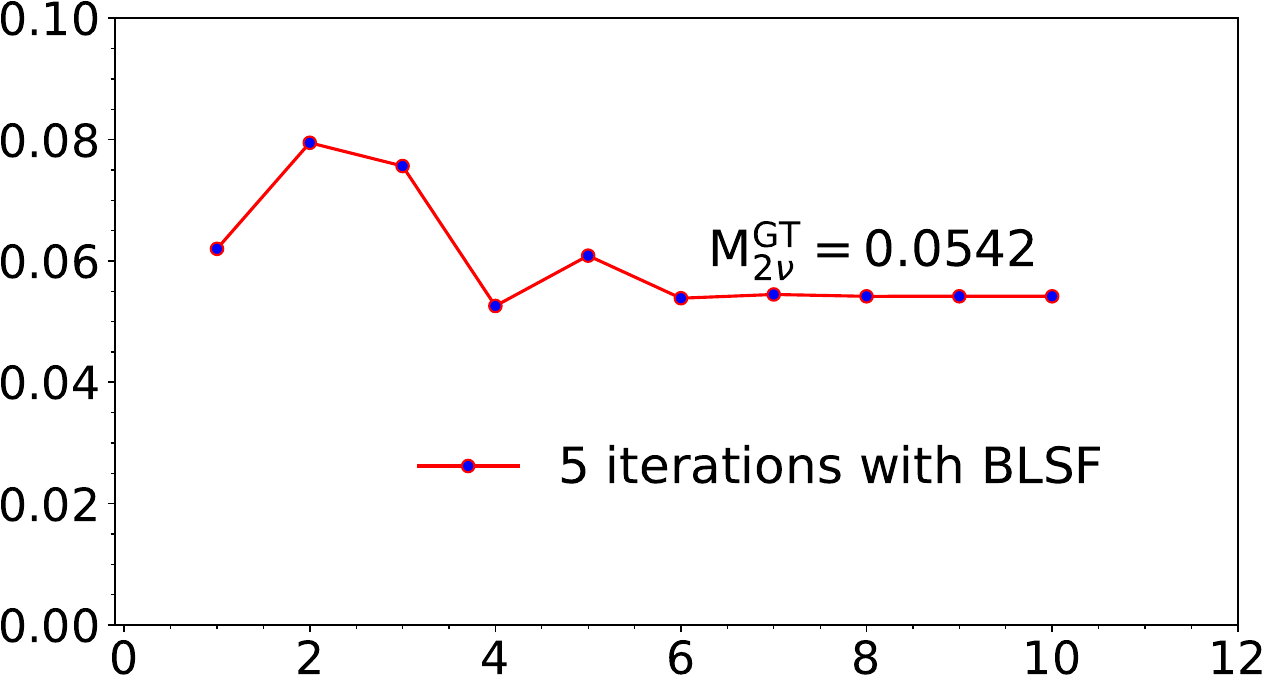}
		\includegraphics[width=50.0mm]{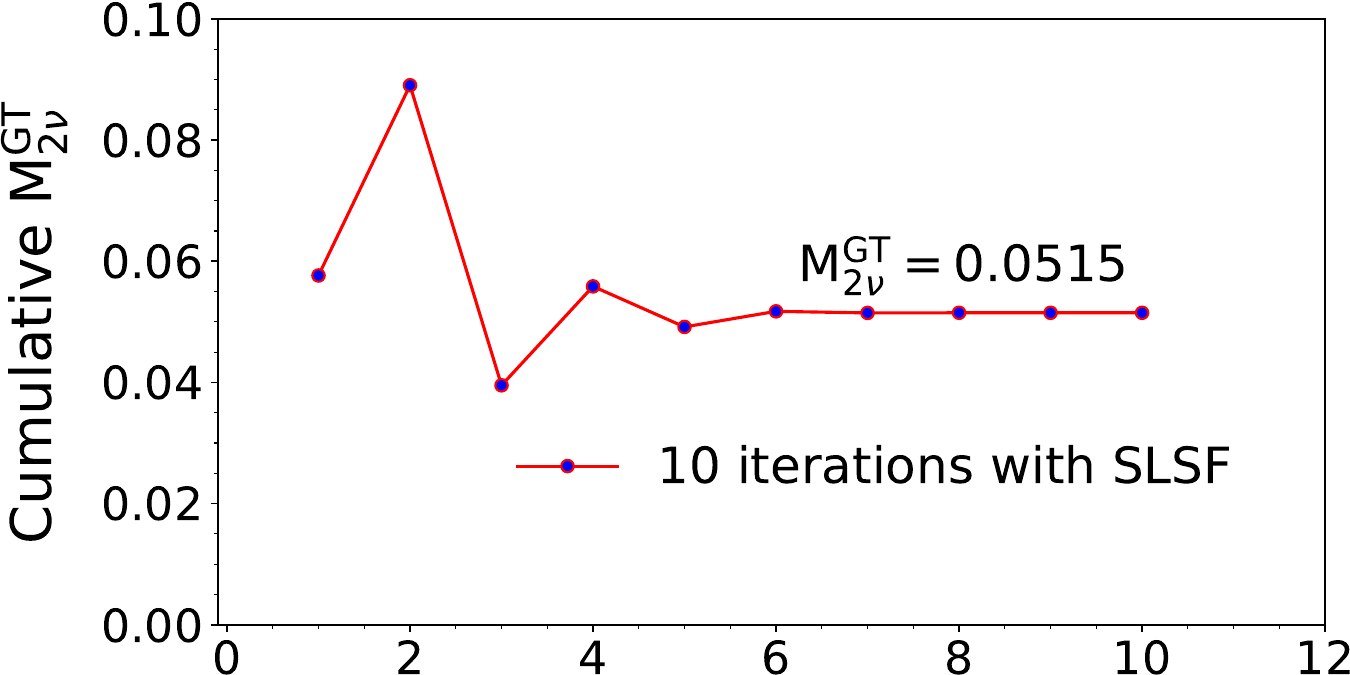}
		\includegraphics[width=46.8mm]{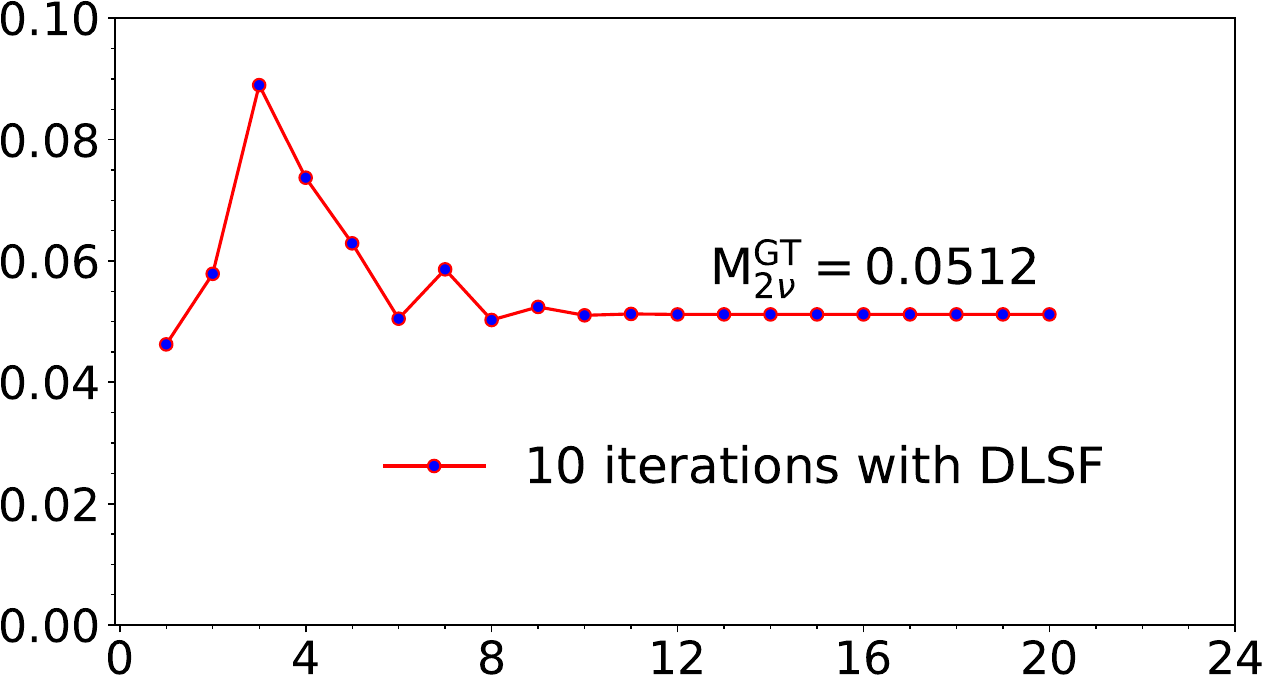}
		\includegraphics[width=46.8mm]{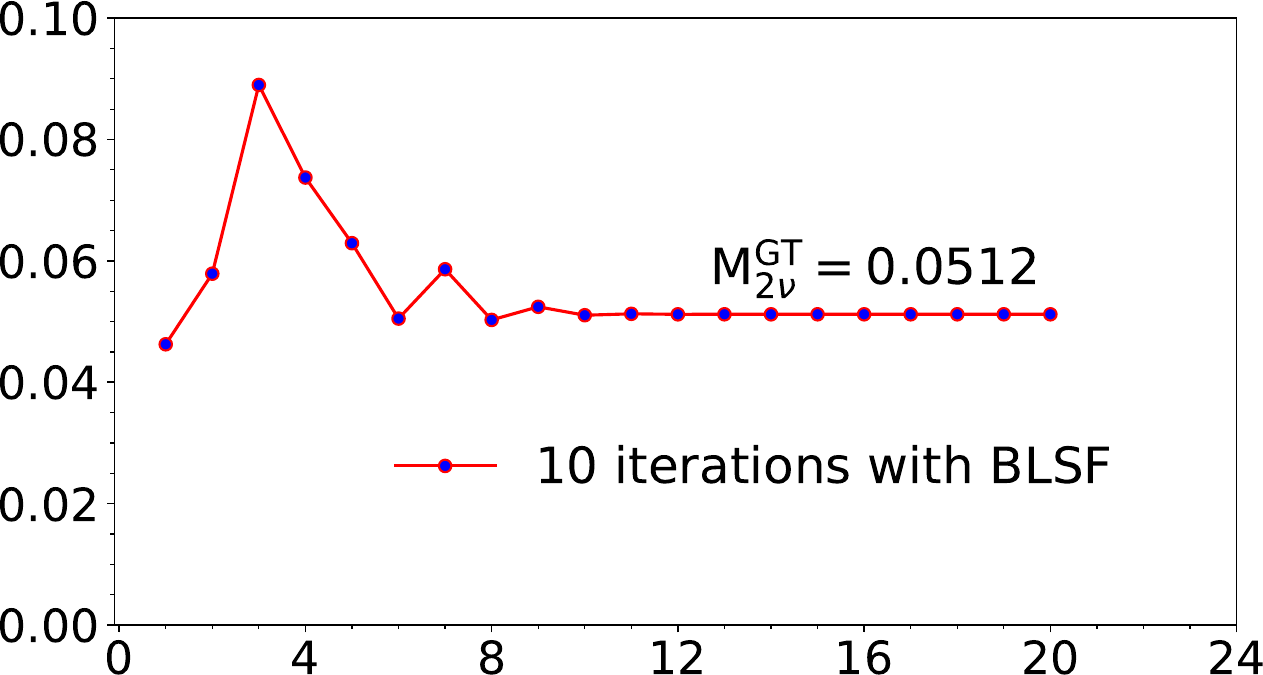}
		\includegraphics[width=50.0mm]{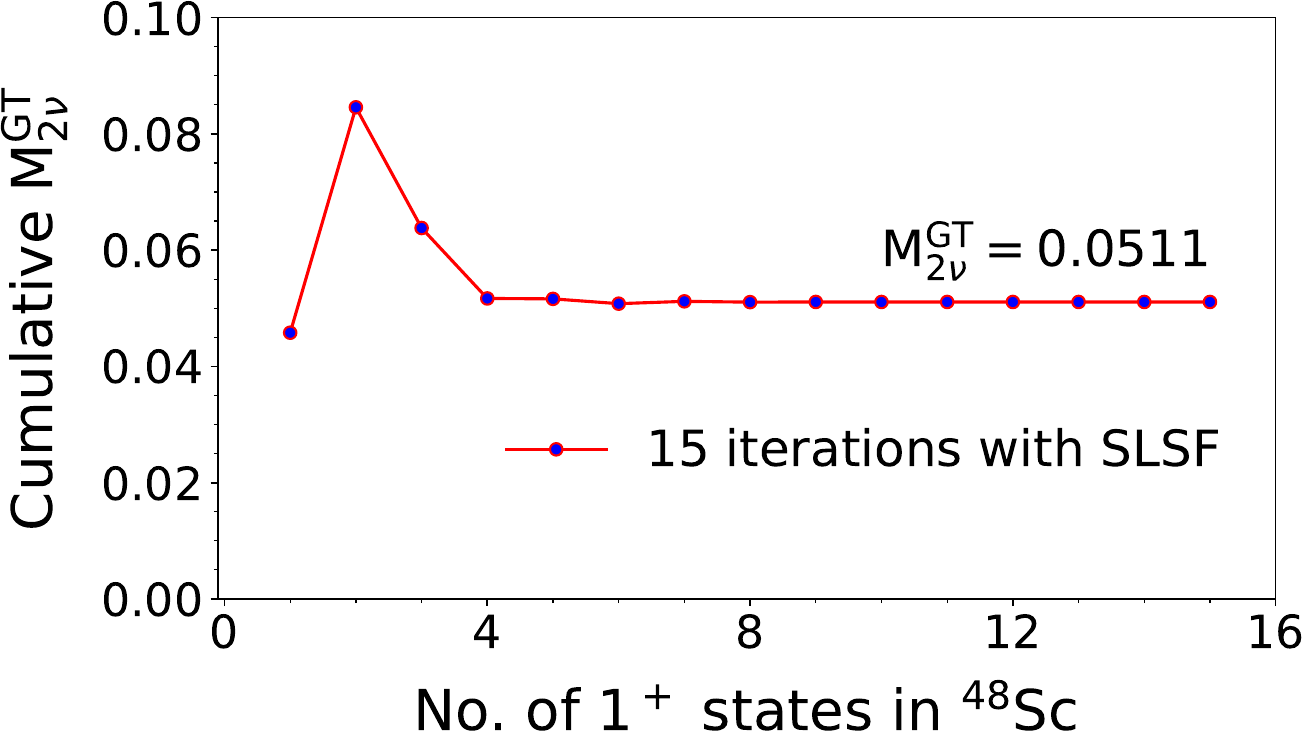}
		\includegraphics[width=46.5mm]{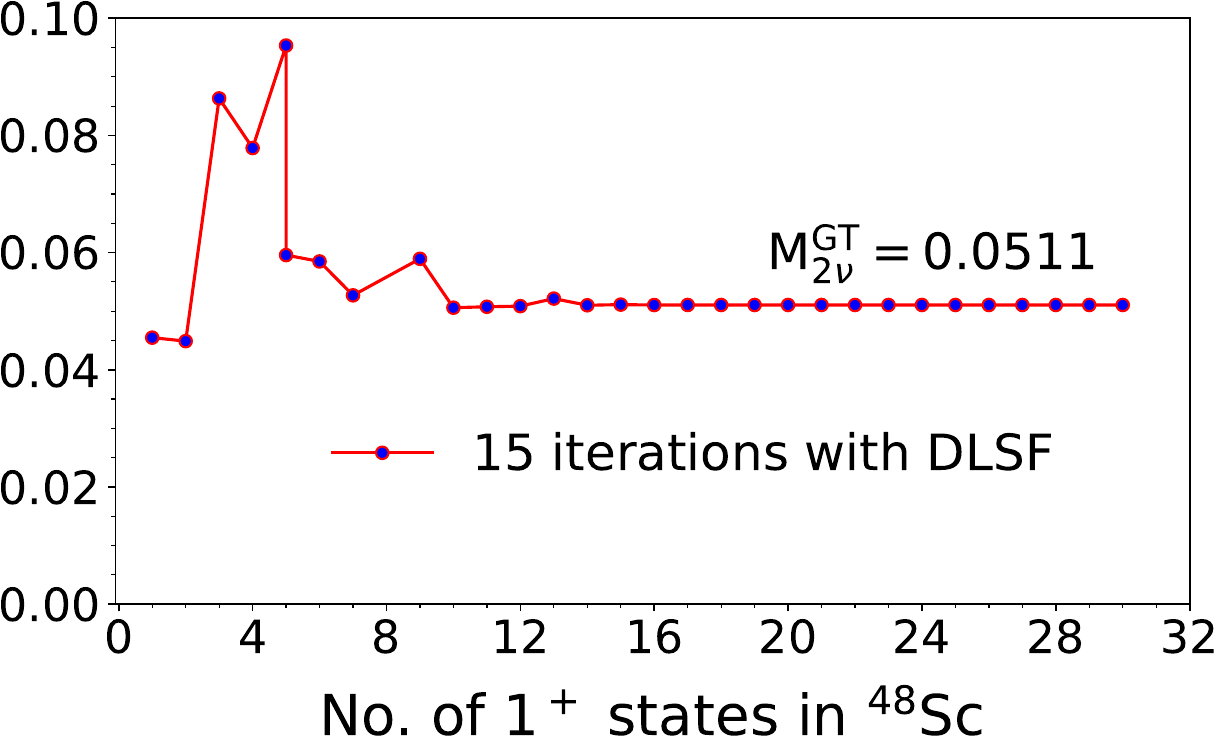}
		\includegraphics[width=46.5mm]{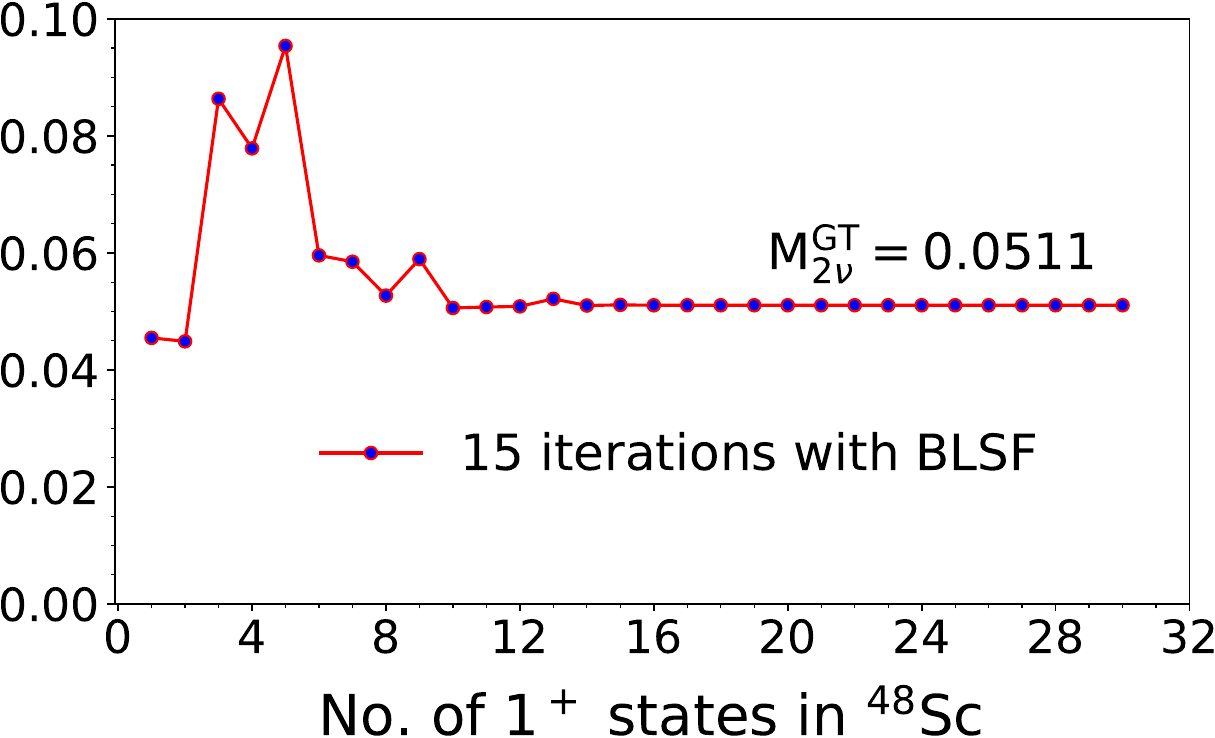}
		\caption{Comparison of the $2\nu\beta\beta$ NME for $^{48}$Ca with the GXPF1A interaction using the SLSF (left), DLSF (middle) and BLSF (right) methods as a function of the number of intermediate $1^+$ states.}
		\label{NME_48Ca_gxpf1a_with_states}
	\end{figure*}

	Here, we test and validate both the DLSF and BLSF methods for $^{48}$Ca using the GXPF1A effective interaction \cite{Honma_gxpf1a} in the full $fp$ shell. We also compare the results with SLSF method. Figure \ref{NME_48Ca_gxpf1a} shows the variation of the cumulative $2\nu\beta\beta$ NME calculated with SLSF, DLSF and BLSF, together with the result from exact diagonalization. The $Q_{\beta\beta}$ values (total energy released in the process) used in the NME calculations are taken from Ref.~\cite{NNDC_Qcal}. The experimental excitation energy of the $1^+_1$ state, 2.2 MeV \cite{NNDC_ENSDF}, is used to determine the common shift applied to the $1^+$ state energies. Using SLSF, we performed 30 iterations for the $^{48}$Ca$(0^+_{\rm g.s.})\rightarrow{}^{48}$Sc$(1^+)$ GT transition, generating 30 intermediate $1^+$ states from one side and obtaining a $2\nu\beta\beta$ NME of 0.0511 (see Fig.~\ref{NME_48Ca_gxpf1a}(a)). In the DSLF calculation, by contrast, 30 iterations were performed separately for the $^{48}$Ca$(0^+_{\rm g.s.})\rightarrow{}^{48}$Sc$(1^+)$ and $^{48}$Ti$(0^+_{\rm g.s.})\rightarrow{}^{48}$Sc$(1^+)$ GT transitions, generating 30 $1^+$ states from each side. After L\"owdin orthogonalization of the combined 60 intermediate $1^+$ states and subsequent rediagonalization of the Hamiltonian in the resulting subspace, 59 Ritz vectors are obtained. Using these 59 intermediate $1^+$ states, the calculated $2\nu\beta\beta$ NME converges to a value of 0.0511. Similarly, in the BLSF method, 30 iterations generate 60 intermediate $1^+$ states. The resulting $2\nu\beta\beta$ NME again converges to 0.0511. Finally, in the exact-diagonalization calculation, we include the lowest 2500 $1^+$ eigenstates of $^{48}$Sc, which yields the same saturated NME as SLSF, DLSF and BLSF approaches (see Fig. \ref{NME_48Ca_gxpf1a}(d)). Although Ref.~\cite{Kostensalo_48Ca} reports a calculation including the maximum possible 9470 $1^+$ eigenstates of $^{48}$Sc, 2500 states are sufficient to clearly establish the convergence of the $2\nu\beta\beta$ NME. These results demonstrate that only 30 iterations with each of the SLSF, DLSF, and BLSF methods are sufficient to accurately reproduce the converged NME, eliminating the need to explicitly calculate several thousand intermediate $1^+$ eigenstates and substantially reducing the computational cost. For $^{48}$Ca in the full $fp$ shell, exact diagonalization remains computationally manageable, and the $2\nu\beta\beta$ NME is already fully converged within 1000 $1^+$ eigenstates. However, exact diagonalization becomes increasingly challenging in larger model spaces, particularly for medium- and heavy-mass candidates, where a larger number of intermediate $1^+$ eigenstates may also be required for NME convergence. These cases are discussed in more detail in Sec. \ref{sec:level4}.

	\begin{figure*}
		\includegraphics[width=35.83mm]{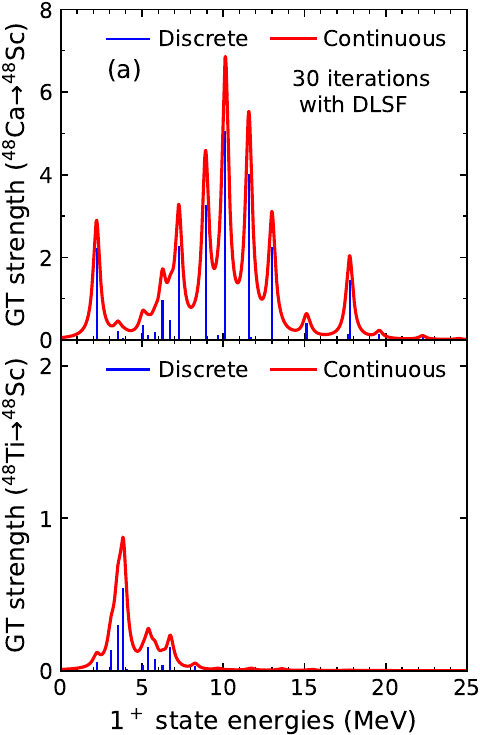}
		\includegraphics[width=33.55mm]{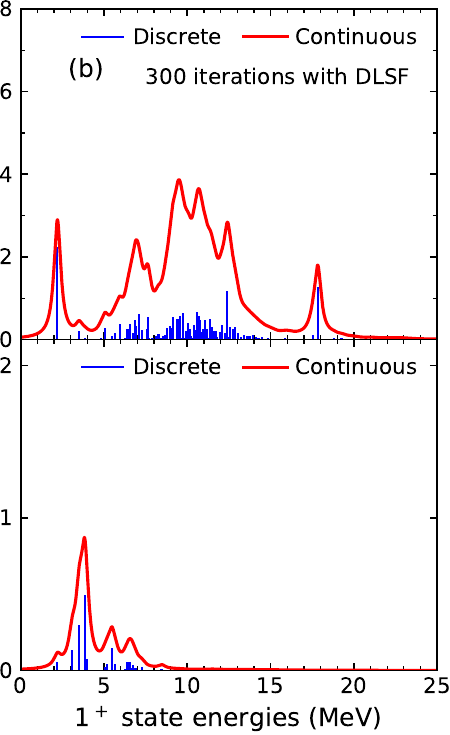}
		\includegraphics[width=33.55mm]{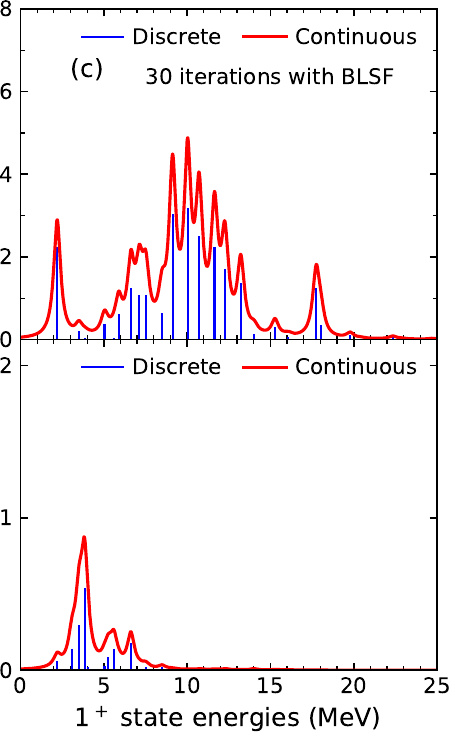}
		\includegraphics[width=33.55mm]{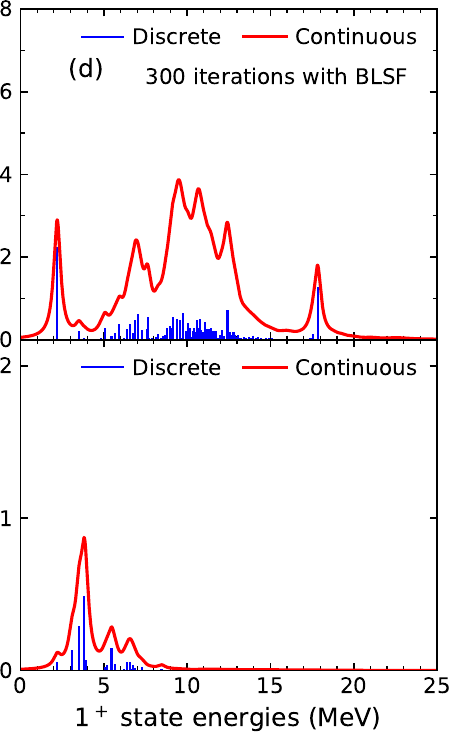}
		\includegraphics[width=33.55mm]{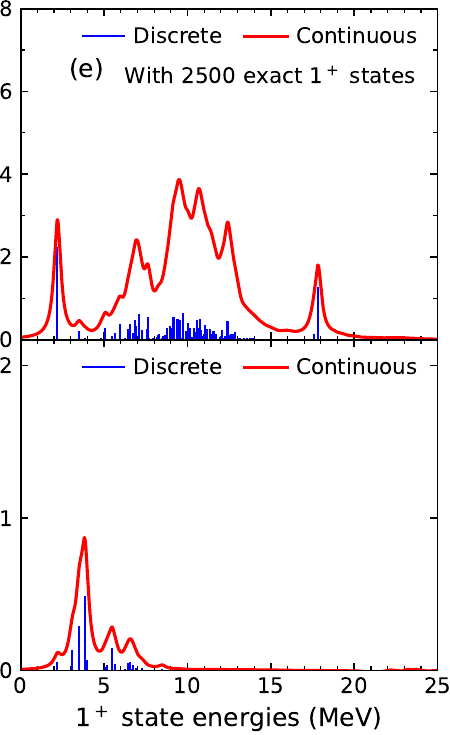}
		\caption{GT strength distributions for the $^{48}$Ca$(0^+_{\rm g.s.})\rightarrow{}^{48}$Sc$(1^+)$ and $^{48}$Ti$(0^+_{\rm g.s.})\rightarrow{}^{48}$Sc$(1^+)$ transitions with GXPF1A interaction using (a) 30 iterations with DLSF, (b) 300 iterations with DLSF, (c) 30 iterations with BLSF, (d) 300 iterations with BLSF, and (e) exact diagonalization with lowest 2500 $1^+$ eigenstates.}
		\label{BGT_48Ca_gxpf1a}
	\end{figure*}

	\begin{table*}
		\caption{Total GT strengths for the $^{48}$Ca$(0^+_{\rm g.s.})\rightarrow{}^{48}$Sc$(1^+)$ and $^{48}$Ti$(0^+_{\rm g.s.})\rightarrow{}^{48}$Sc$(1^+)$ transitions with the GXPF1A interaction, using 30 iterations in SLSF, DLSF, and BLSF methods, along with exact diagonalization including the lowest 2500 $1^+$ eigenstates.}
		\begin{ruledtabular}
			\begin{tabular}{cccccc}
				
				Transition & \multicolumn{5}{c}{Method}	 \\
				
				\cline{2-6}
				
				& SLSF (from $^{48}$Ca) & SLSF (from $^{48}$Ti) & DLSF & BLSF & Exact diagonalization (2500 $1^+$ eigenstates) \\
				
				\hline
				
				$^{48}$Ca$\rightarrow{}^{48}$Sc & 24.000 &  4.183 & 24.000 & 24.000 & 23.988 \\
				
				$^{48}$Ti$\rightarrow{}^{48}$Sc & 0.957 &  1.587 & 1.587 & 1.587 & 1.581 \\
				
			\end{tabular}
		\end{ruledtabular}
		\label{Sum_rule_GT}
	\end{table*}

	To examine the convergence more closely, we perform calculations with 5, 10, and 15 iterations and plot the cumulative NME as a function of the number of intermediate $1^+$ states in Fig.~\ref{NME_48Ca_gxpf1a_with_states}. For each of the three iteration counts, DLSF and BLSF yield the same final NME. With 5 iterations, both methods give an NME of 0.0542; increasing the number of iterations to 10 gives 0.0512, while 15 iterations yield the fully converged value of 0.0511. Thus, DLSF and BLSF exhibit very similar convergence behavior in this case. The SLSF method also reaches the fully converged value within 15 iterations, although its cumulative NME shows a somewhat different convergence pattern at 5 and 10 iterations.

	Figure~\ref{BGT_48Ca_gxpf1a} shows the GT strength distributions for the $^{48}$Ca$(0^+_{\rm g.s.})\rightarrow{}^{48}$Sc$(1^+)$  and $^{48}$Ti$(0^+_{\rm g.s.})\rightarrow{}^{48}$Sc$(1^+)$ transitions calculated with the GXPF1A interaction using DLSF and BLSF. The results are compared with those obtained by exact diagonalization including the lowest 2500 intermediate $1^+$ eigenstates. The discrete strengths $S(E_k)$ (strength per state) are indicated by vertical bars, while the corresponding continuous distributions $S_{\rm cont}(E)$ (strength density in MeV$^{-1}$) are obtained by Lorentzian folding of the discrete strengths with a width of $\Gamma=0.5$ MeV and can be expressed as
	\begin{equation}
		S_{\rm cont}(E) = \frac{\Gamma}{2\pi} \sum_{k=1}^n \frac{S(E_k)}{(E - E_k)^2 + (\Gamma/2)^2},
		\label{S_cont}
	\end{equation}
	where $\Gamma$ denotes the full width at half maximum (FWHM) and $E_k$ is the excitation energy of the $k$-th intermediate $1^+$ state. We examine the distribution of the discrete GT strengths, particularly for the $^{48}$Ca$(0^+_{\rm g.s.})\rightarrow{}^{48}$Sc$(1^+)$ transition, since the strengths for the $^{48}$Ti$(0^+_{\rm g.s.})\rightarrow{}^{48}$Sc$(1^+)$ transitions are very small. For 30 iterations, the discrete strengths obtained with both the DLSF and BLSF methods (Figs.~\ref{BGT_48Ca_gxpf1a}(a) and \ref{BGT_48Ca_gxpf1a}(c)) are less fragmented and comparatively larger. With increasing iterations, which produce a larger number of intermediate $1^+$ states, the discrete strengths become more fragmented and smaller (Figs.~\ref{BGT_48Ca_gxpf1a}(b) and \ref{BGT_48Ca_gxpf1a}(d)), while conserving the total strength. We also note that, with 300 iterations in both the DLSF and BLSF methods, the discrete and continuous strength distributions approach those obtained from exact diagonalization (see Fig. \ref{BGT_48Ca_gxpf1a}(e)). Furthermore, the smooth continuous strength distribution for a larger number of iterations indicates that the overall distribution becomes stable after a given number of iterations. The GT strength distribution for the $^{48}$Ca$(0^+_{\rm g.s.})\rightarrow{}^{48}$Sc$(1^+)$ transition has a giant-resonance peak around 10 MeV, which is reproduced by both the DLSF and BLSF methods. However, the GT strength from the $^{48}$Ti$(0^+_{\rm g.s.})\rightarrow{}^{48}$Sc$(1^+)$ transition is very small in this energy region, suppressing the corresponding contributions to the $2\nu\beta\beta$ NME. Consequently, the cumulative NME exhibits only minor fluctuations around the giant-resonance energy. These strength distributions demonstrate that both the DLSF and BLSF methods capture all the GT strength allowed by the adopted model space.

	As shown in Fig.~\ref{NME_48Ca_gxpf1a_with_states}, all three methods yield a converged $2\nu\beta\beta$ NME within 15 iterations. In particular, SLSF reproduces the converged NME regardless of whether its intermediate $1^+$ states are generated from the $^{48}$Ca or $^{48}$Ti ground state. Nevertheless, neither set of SLSF states simultaneously reproduces the total GT strengths for both transitions. Table \ref{Sum_rule_GT} summarizes the total GT strengths obtained with the SLSF method using 30 iterations from either the $^{48}$Ca or $^{48}$Ti side, the DLSF and BLSF methods using 30 iterations, and exact diagonalization including the lowest 2500 $1^+$ eigenstates. In the SLSF calculations, the resulting intermediate states reproduce the total GT strength only for the transition used to generate them. Specifically, states generated from the $^{48}$Ca ground state reproduce the total strength for the $^{48}$Ca$(0^+_{\rm g.s.})\rightarrow{}^{48}$Sc$(1^+)$ transition but not that for the $^{48}$Ti$(0^+_{\rm g.s.})\rightarrow{}^{48}$Sc$(1^+)$ transition, and vice versa for states generated from the $^{48}$Ti ground state. In contrast, both the DLSF and BLSF methods preserve the total GT strengths for both transitions (see Table \ref{Sum_rule_GT}). For the exact diagonalization calculation with the lowest 2500 $1^+$ eigenstates, the corresponding total GT strengths are underestimated by not more than $0.4\%$ for both the $^{48}$Ca$(0^+_{\rm g.s.})\rightarrow{}^{48}$Sc$(1^+)$ and $^{48}$Ti$(0^+_{\rm g.s.})\rightarrow{}^{48}$Sc$(1^+)$ transitions and can be recovered by including a few hundred additional $1^+$ eigenstates. While all three methods yield similarly rapid convergence of the $2\nu\beta\beta$ NME in this benchmark, DLSF and BLSF have the additional advantage of simultaneously reproducing the total GT strengths associated with both the initial and final nuclei.

	\begin{figure*}[t]
		\centering
		\includegraphics[width=45.0mm]{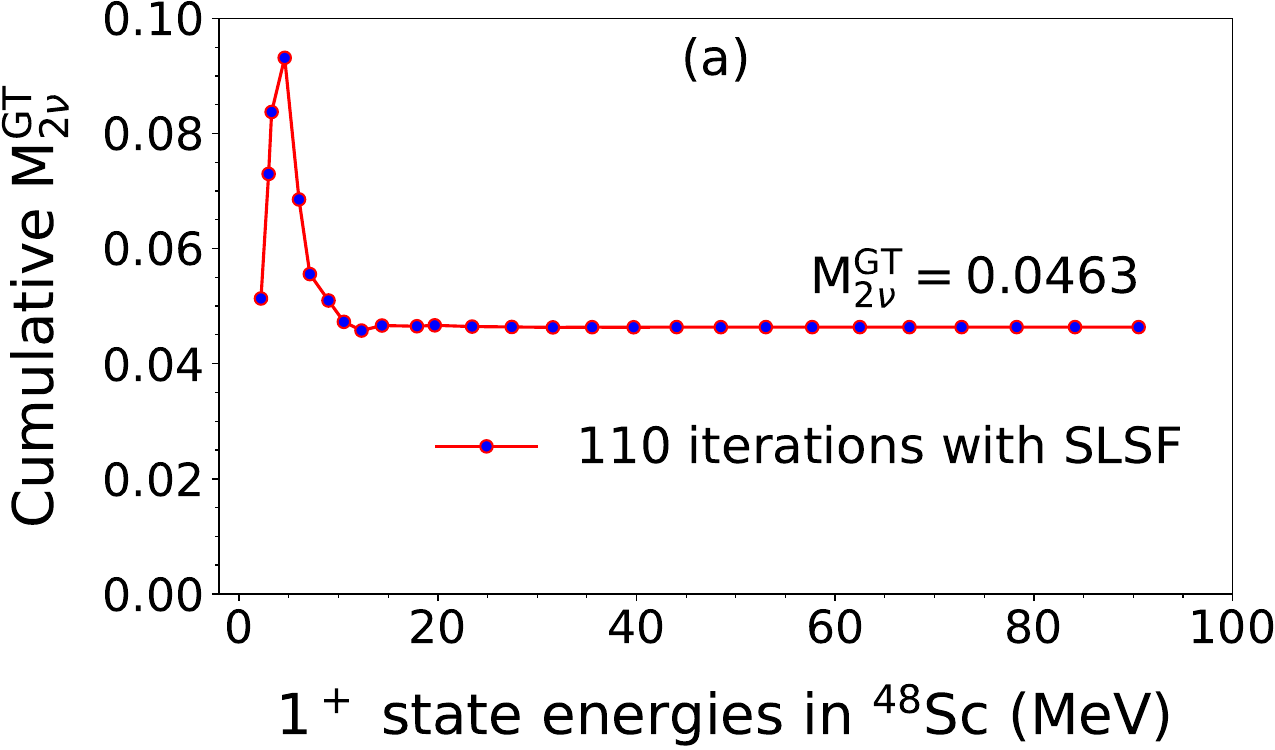}
		\includegraphics[width=41.8mm]{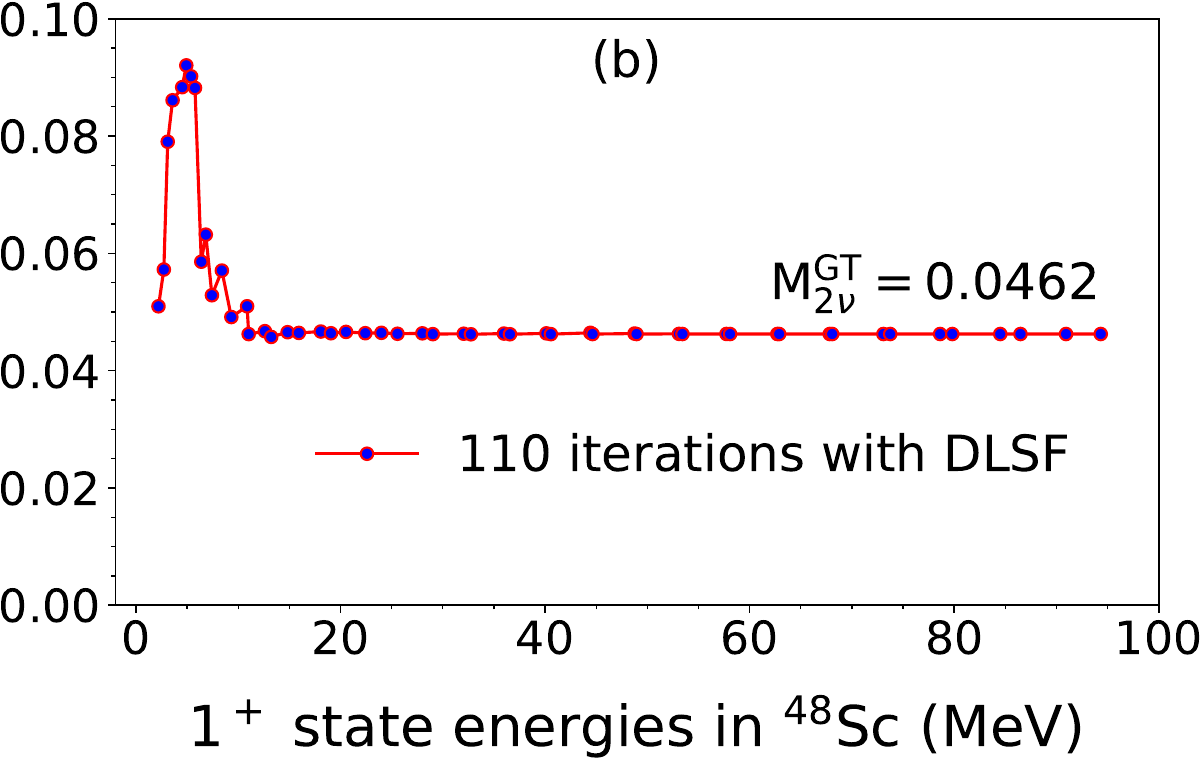}
		\includegraphics[width=41.8mm]{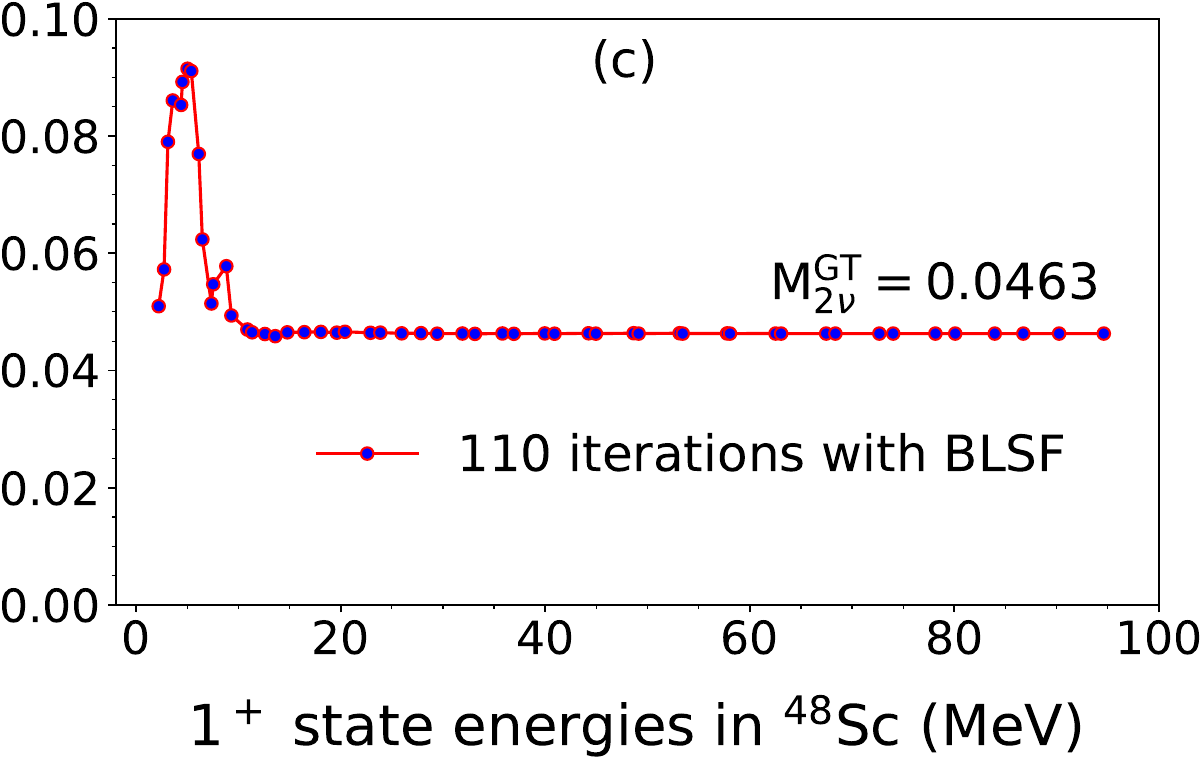}
		\includegraphics[width=41.8mm]{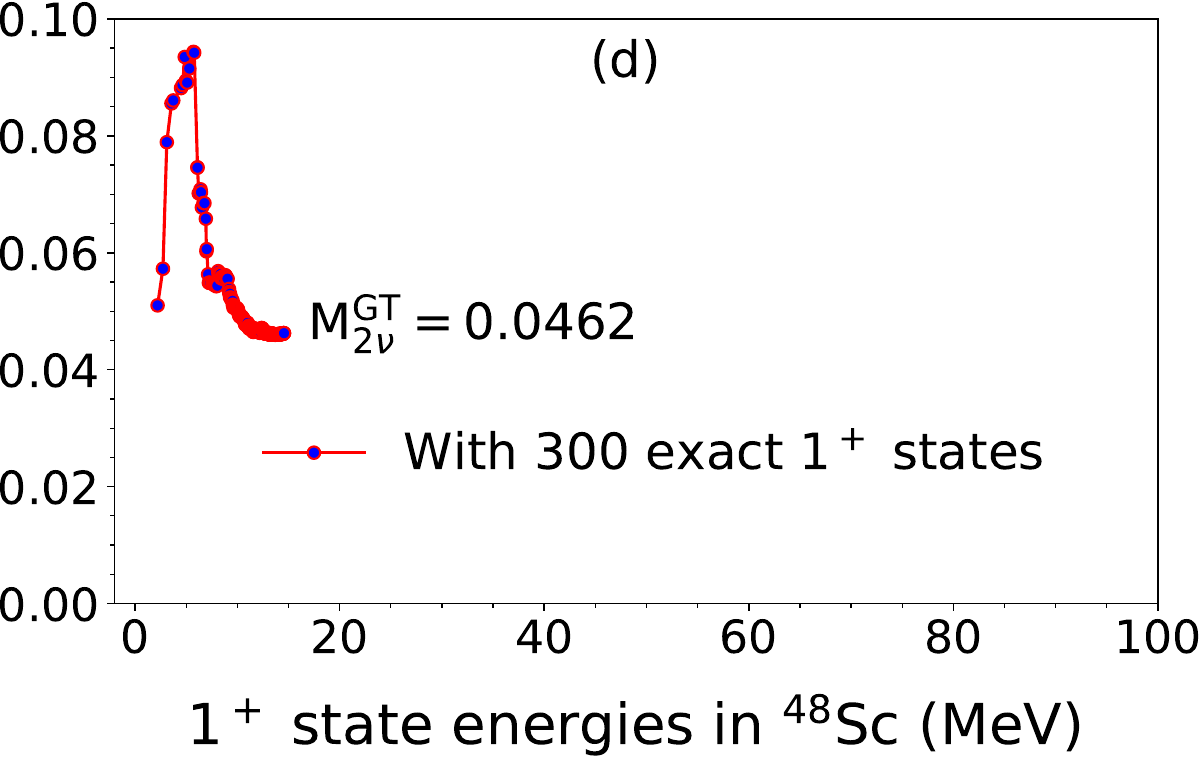}
		\caption{Comparison of the $2\nu\beta\beta$ NME for $^{48}$Ca with the SDPF-MU interaction obtained using the (a) SLSF, (b) DLSF, (c) BLSF, and (d) exact-diagonalization methods.}
		\label{NME_48Ca_sdpfmu}
		
		\vspace{0.44cm}
		
		\includegraphics[width=45.0mm]{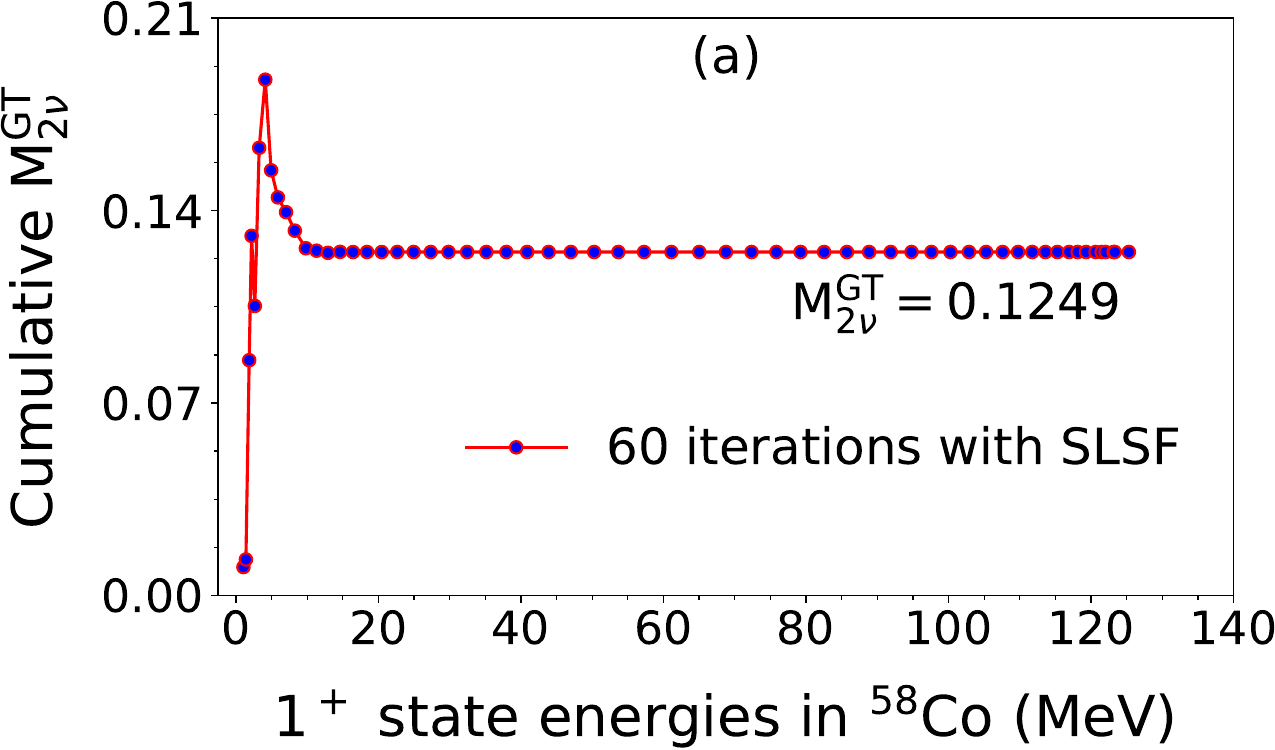}
		\includegraphics[width=41.8mm]{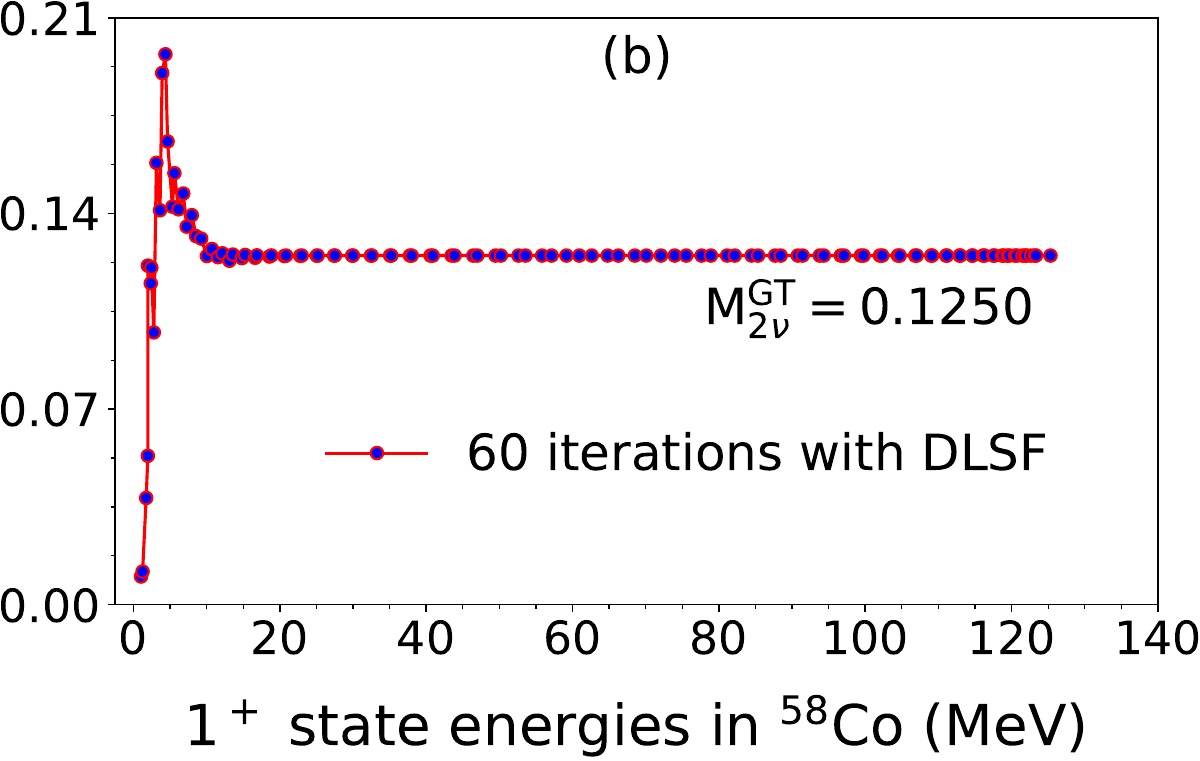}
		\includegraphics[width=41.8mm]{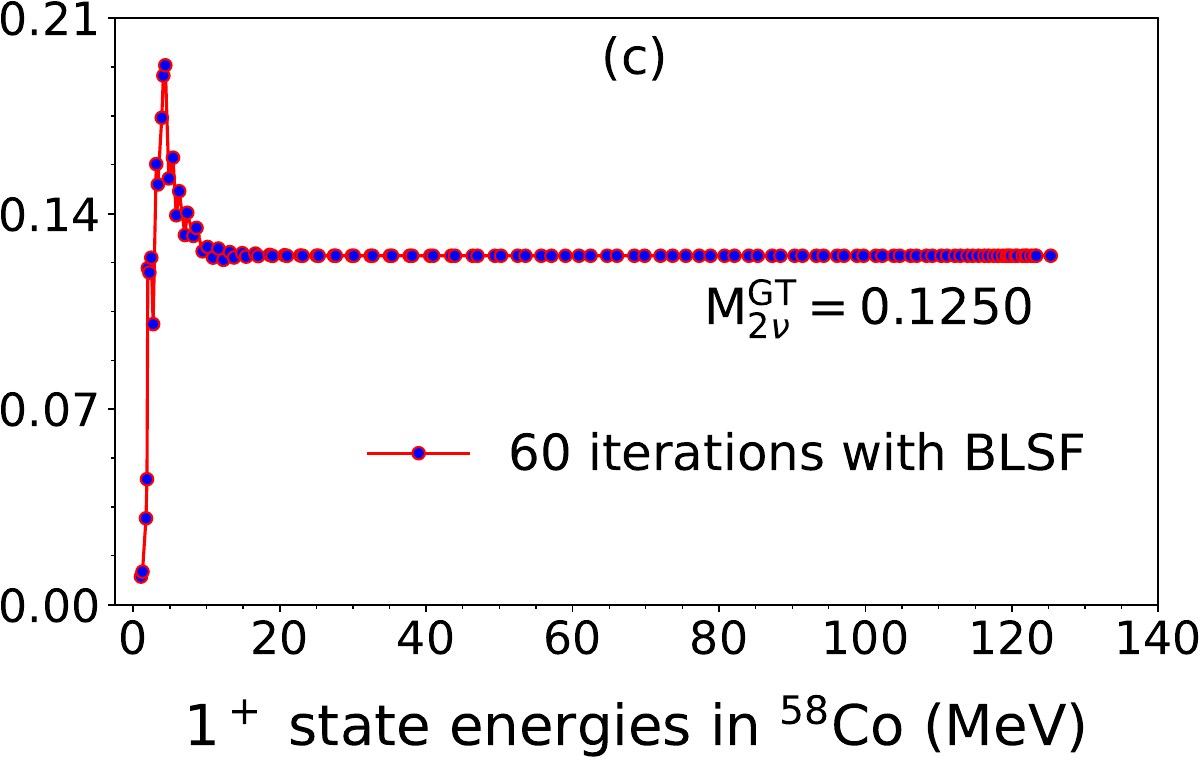}
		\includegraphics[width=41.8mm]{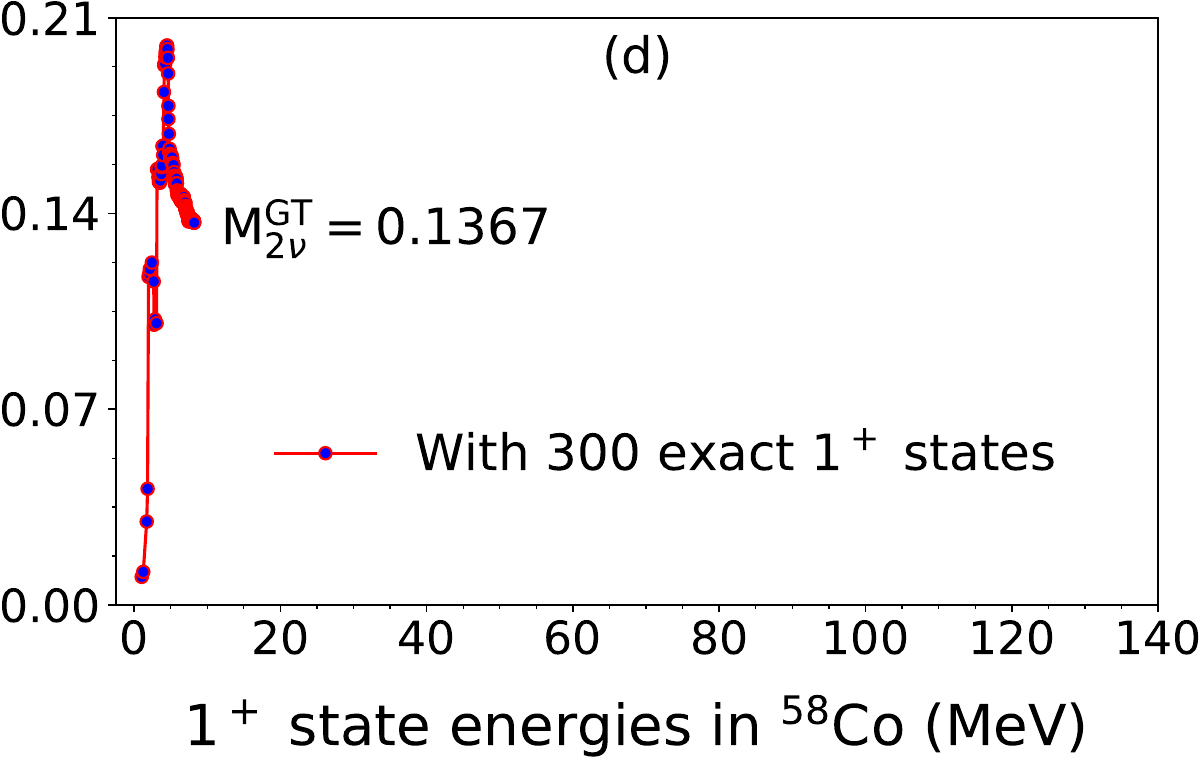}
		\caption{Comparison of the $2\nu$ECEC NME for $^{58}$Ni with the GXPF1A interaction obtained using the (a) SLSF, (b) DLSF, (c) BLSF, and (d) exact-diagonalization methods.}
		\label{NME_58Ni_gxpf1a}
		
		\vspace{0.44cm}
		
		\includegraphics[width=45.0mm]{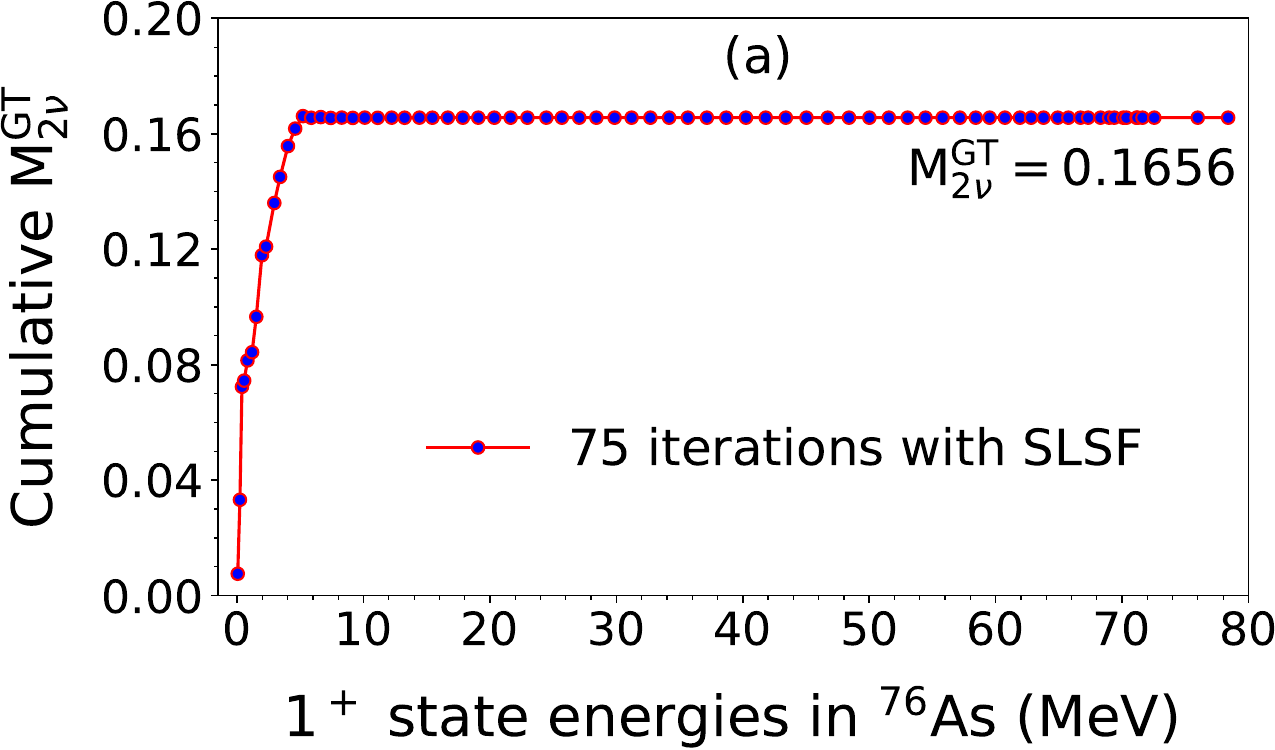}
		\includegraphics[width=41.8mm]{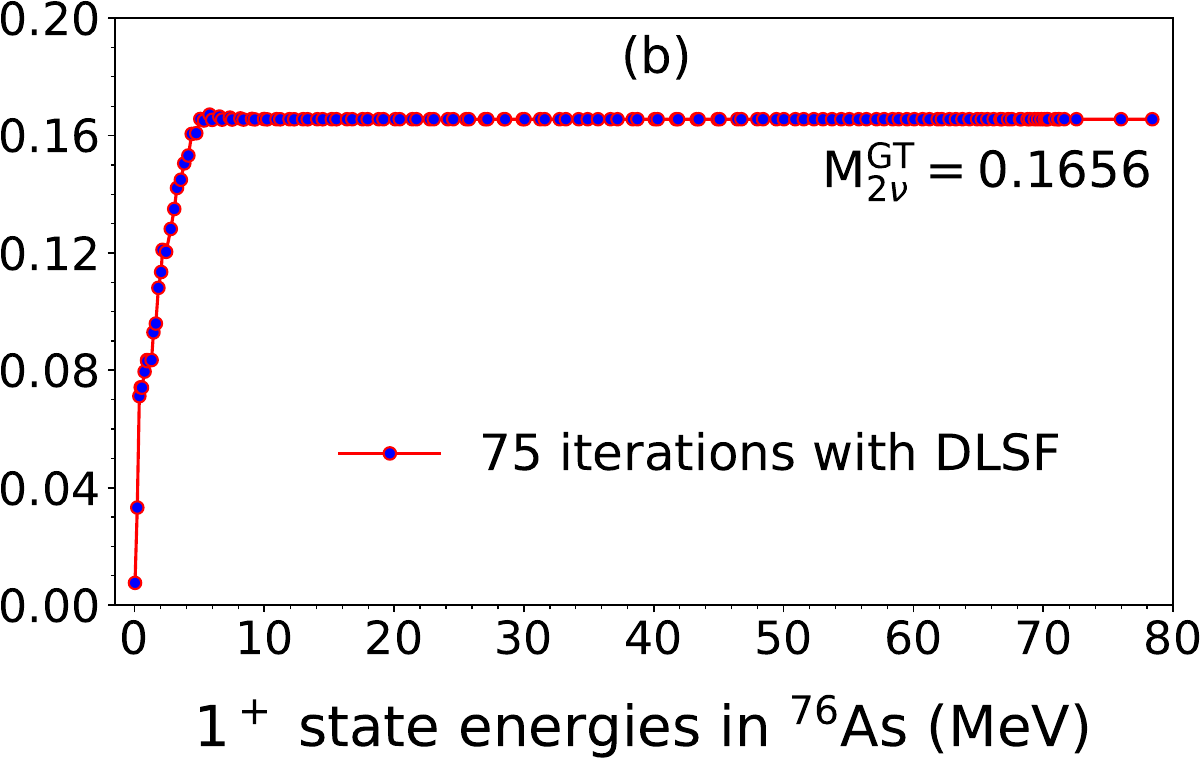}
		\includegraphics[width=41.8mm]{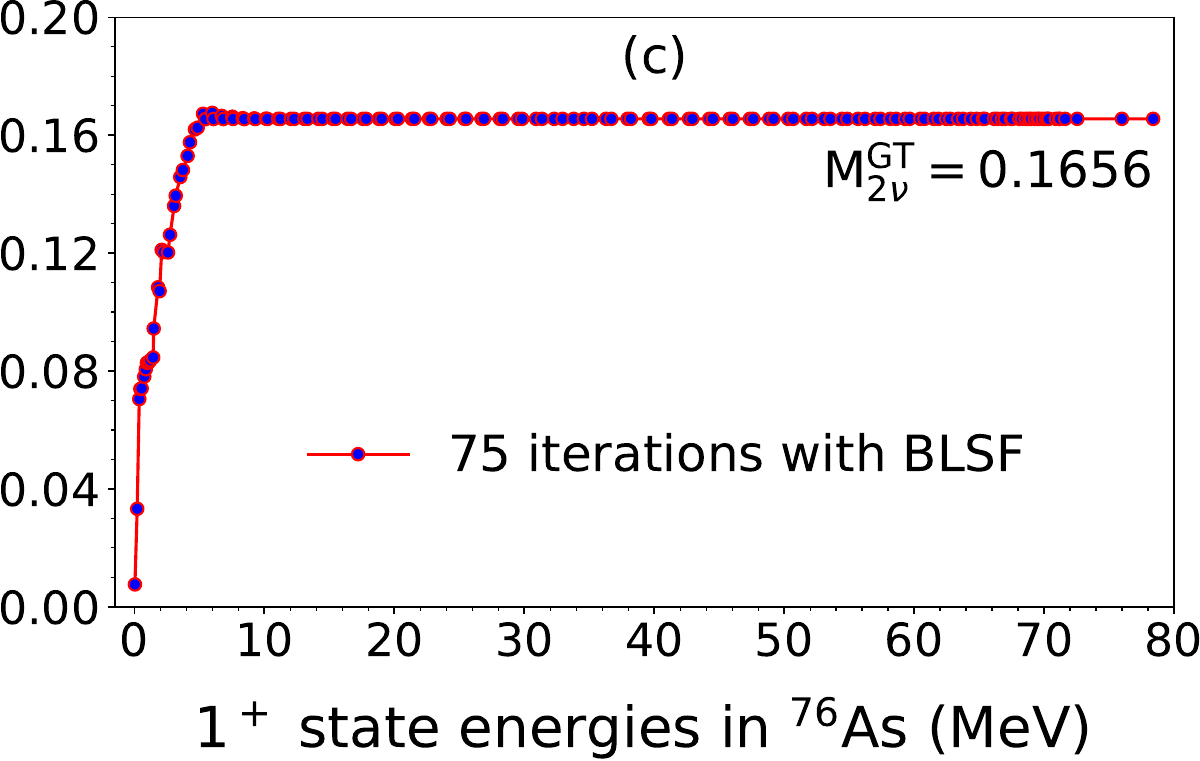}
		\includegraphics[width=41.8mm]{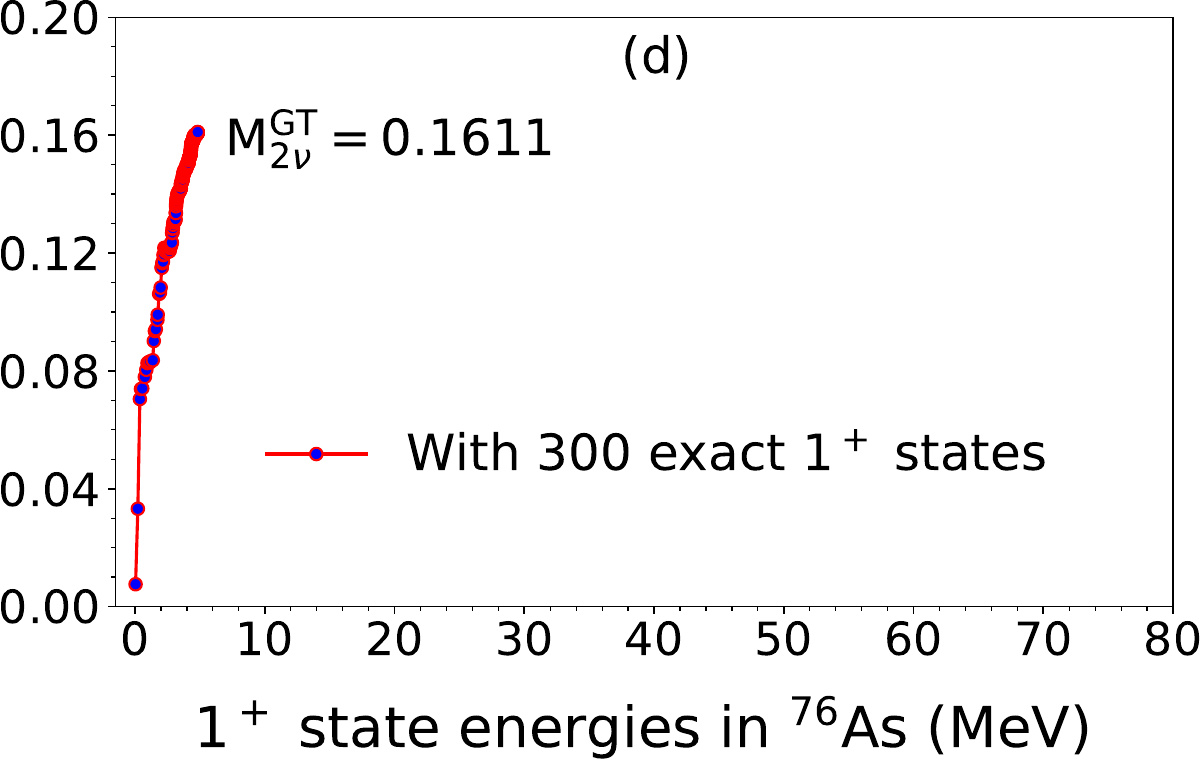}
		\caption{Comparison of the $2\nu\beta\beta$ NME for $^{76}$Ge with the JUN45 interaction obtained using the (a) SLSF, (b) DLSF, (c) BLSF, and (d) exact-diagonalization methods.}
		\label{NME_76Ge_jun45}
		
		\vspace{0.44cm}
		
		\includegraphics[width=45.0mm]{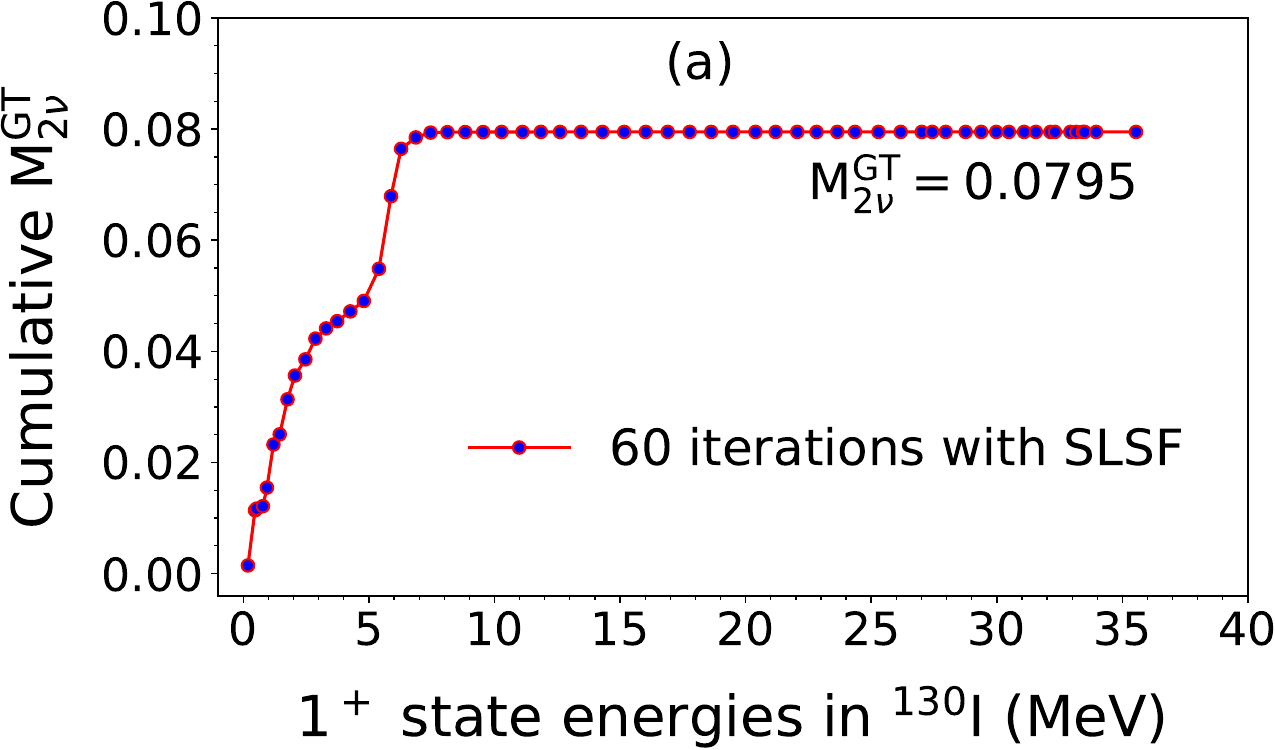}
		\includegraphics[width=41.8mm]{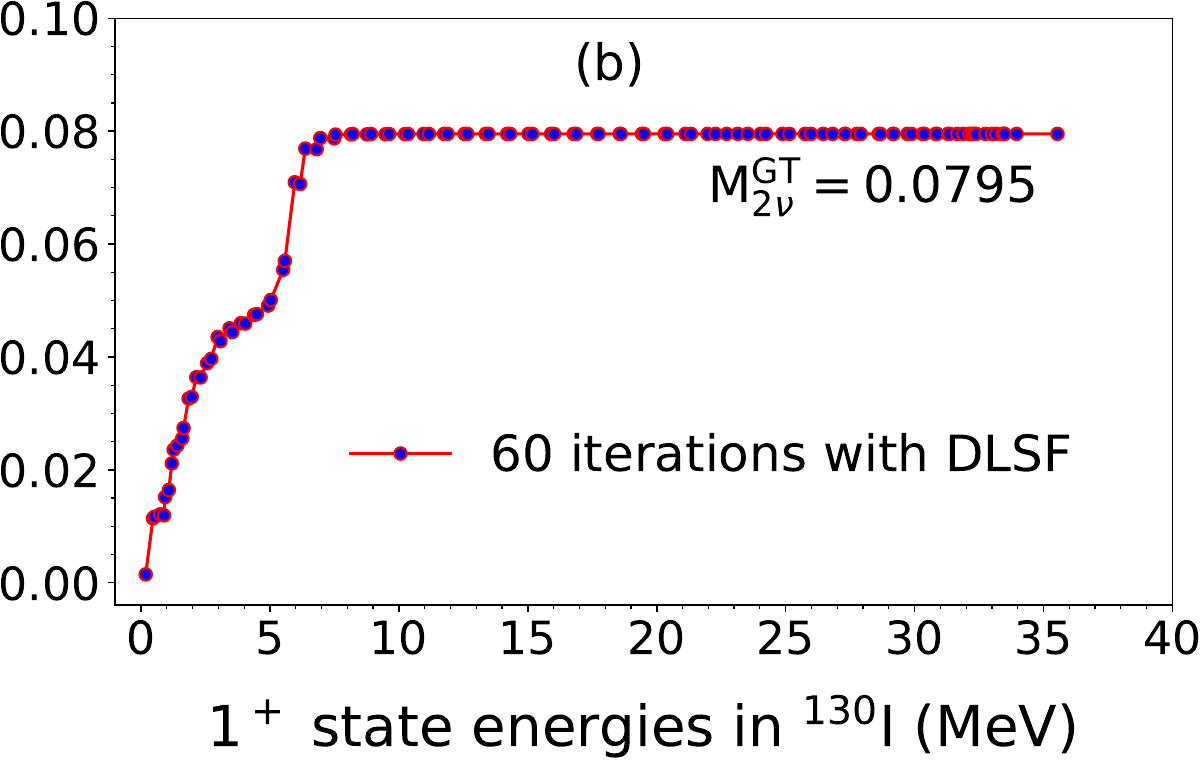}
		\includegraphics[width=41.8mm]{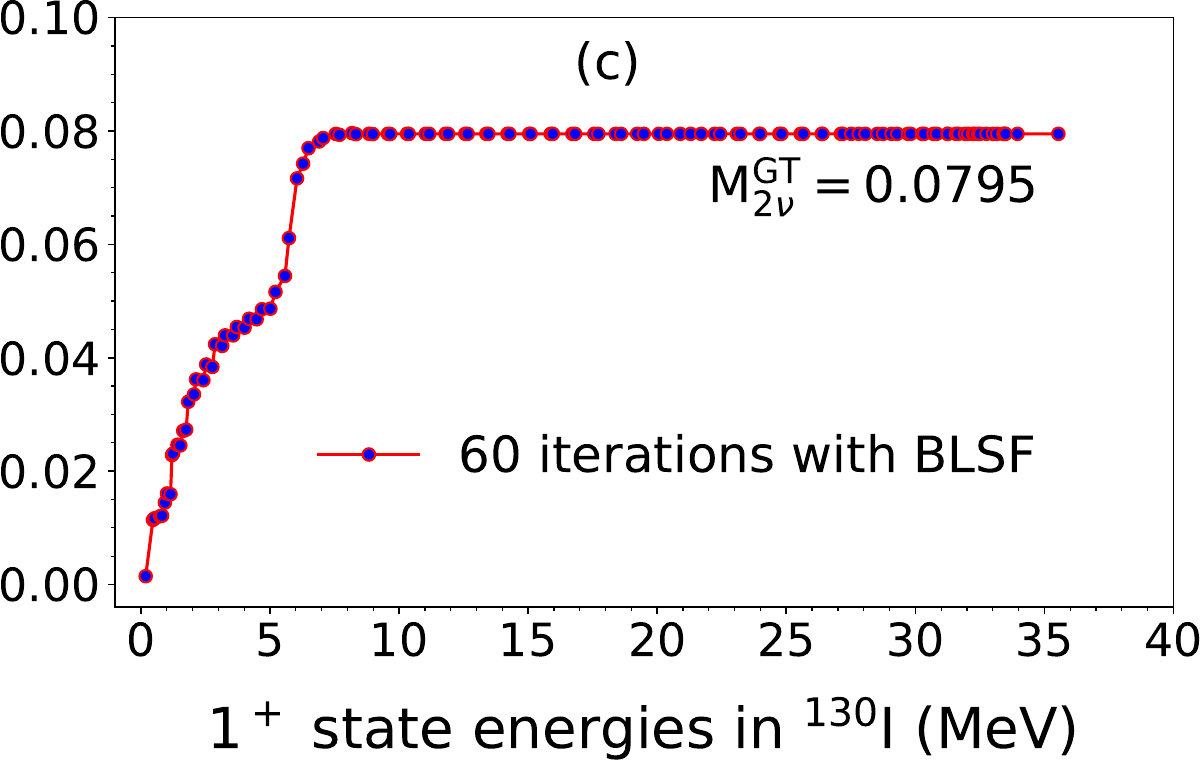}
		\includegraphics[width=41.8mm]{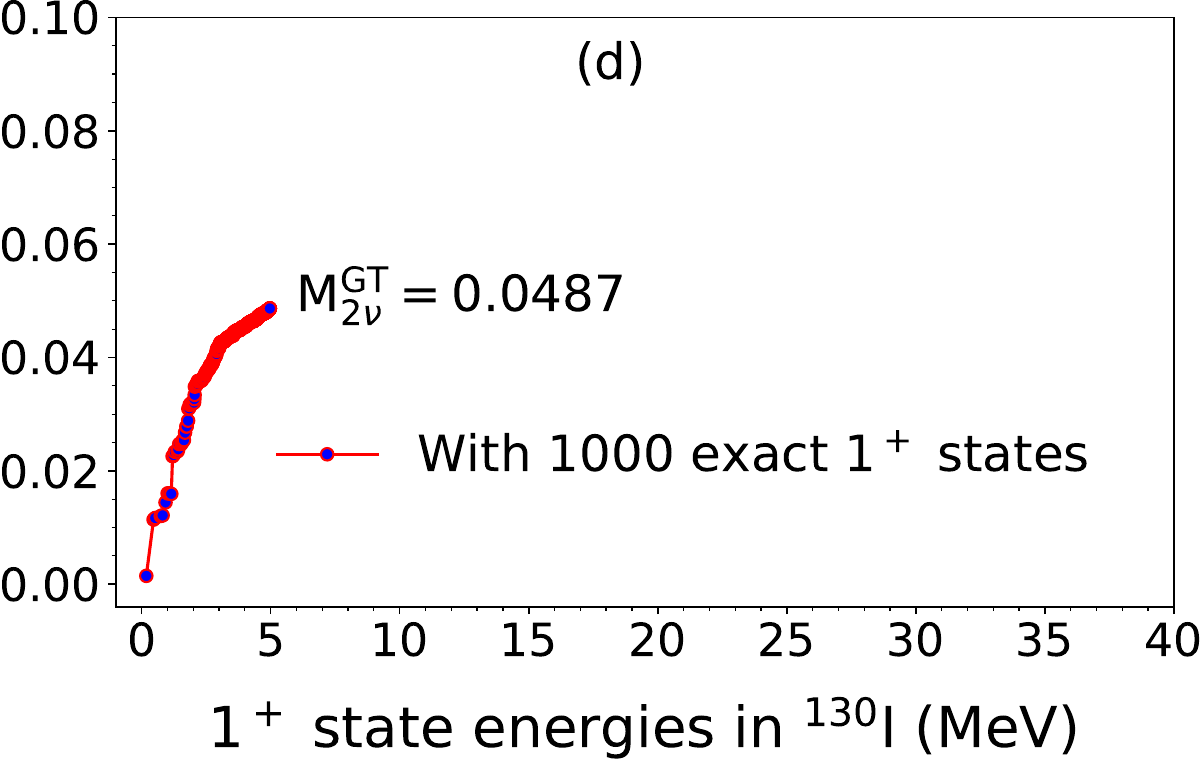}
		\caption{Comparison of the $2\nu\beta\beta$ NME for $^{130}$Te with the SN100PN interaction obtained using the (a) SLSF, (b) DLSF, (c) BLSF, and (d) exact-diagonalization methods.}
		\label{NME_130Te_sn100pn}
	\end{figure*}

	\section{\label{sec:level4}Application to larger systems in different mass regions}
	
	Here, we apply the SLSF, DLSF, and BLSF methods to $2\nu\beta\beta$ candidates in several mass regions: $^{48}$Ca in the $sdpf$ shell, $^{76}$Ge in the $f_{5/2}pg_{9/2}$ shell, and $^{130}$Te in the $g_{7/2}dsh_{11/2}$ shell. Because obtaining enough exact intermediate $1^+$ eigenstates to fully converge the NMEs is computationally challenging in these large-scale calculations, comparisons with exact diagonalization are made only within the computationally tractable range. We also present a shell-model calculation of the $2\nu$ECEC NME for $^{58}$Ni in the $fp$ shell, which, to our knowledge, has not previously been reported.

	We first consider $^{48}$Ca with the SDPF-MU interaction \cite{sdpf-mu}, using the $sd$ and $pf$ harmonic-oscillator shells and allowing up to $2\hbar\omega$ $sdpf$ cross-shell excitations. To suppress spurious center-of-mass (c.m.) excitations, we employed the Lawson prescription \cite{Gloeckner}, adding the c.m. Hamiltonian $H_{\rm cm}$ to the shell-model Hamiltonian $H_{\rm SM}$ as
	\begin{equation}
		H=H_{\rm SM}+\beta_{\rm cm}H_{\rm cm},
	\end{equation}
	where
	\begin{equation}
		H_{\rm cm}=\frac{\boldsymbol{P}_{\rm cm}^{2}}{2Am}
		+\frac{1}{2}Am\omega^{2}\boldsymbol{R}_{\rm cm}^{2}-\frac{3}{2}\hbar\omega,
	\end{equation}    
	with $\boldsymbol{R}_{\rm cm}=\frac{1}{A}\sum_{i=1}^{A}\boldsymbol{r}_i$ and $\boldsymbol{P}_{\rm cm}=\sum_{i=1}^{A}\boldsymbol{p}_i$ denoting the c.m. position and momentum operators, respectively; $\boldsymbol{r}_i$ and $\boldsymbol{p}_i$ are the position and momentum operators of the $i$-th nucleon. Here, $A$ is the mass number, $m$ is the nucleon mass, and $\omega$ is the harmonic-oscillator frequency. The dimensionless parameter $\beta_{\rm cm}$, which controls the strength of the c.m. Hamiltonian contribution, is set to 10. For the SLSF, DLSF, and BLSF methods, 110 iterations were performed, by which the resulting $2\nu\beta\beta$ NME had completely converged. Some of the higher-lying Ritz states obtained with the three Lanczos-based methods contain substantial c.m. contamination, as indicated by their large expectation values of $H_{\rm cm}$. We therefore exclude states with $\langle H_{\rm cm}\rangle >$ 0.1 MeV from the NME calculations. In each approach (SLSF, DLSF and BLSF), the $2\nu\beta\beta$ NME saturates at approximately 0.0463 (see Fig. \ref{NME_48Ca_sdpfmu}). Here, we again used the experimental excitation energy of the $1^+_1$ state, 2.2 MeV \cite{NNDC_ENSDF}, to determine the common shift applied to the $1^+$ state energies. With our current computational resources, exact diagonalization of the lowest 300 $1^+$ eigenstates of $^{48}$Sc is still feasible. Remarkably, these 300 states are already sufficient to yield an NME of 0.0462, in close agreement with the converged values obtained using SLSF, DLSF, and BLSF. Also, since these 300 $1^+$ eigenstates span excitation energies up to 15 MeV, Fig. \ref{NME_48Ca_sdpfmu} shows that the cumulative NMEs from SLSF, DLSF and BLSF methods exhibit only negligible fluctuations above 15 MeV. 
	An experimental estimate of the NME can be obtained as $|M_{2\nu}|^{\rm eff}=0.020\pm0.002$ using Eq.~(\ref{eq1}), together with $G^{2\nu}$ from Ref.~\cite{Neacsu}, the experimental half-life from Ref.~\cite{Arnold}, and  the free-nucleon axial-vector coupling constant ($g_{\rm A}=1.27$). Following Ref. \cite{Iwata}, quenching factors $q=0.74$ and 0.71 can be applied to the calculated NMEs in the $fp$ and $sdpf$ shells, respectively, as $\mathrm{NME}\times q^2$. The resulting NMEs for $^{48}$Ca with the GXPF1A and SDPF-MU interactions become 0.0280 and 0.0233, respectively. Even before applying the quenching factor, enlarging the model space from the $fp$ to the $sdpf$ shell reduces the calculated $2\nu\beta\beta$ NME from 0.0511 to 0.0463, bringing it closer to the experimental estimate. In Ref.~\cite{Iwata}, the same enlargement of the model space was found to increase the calculated $0\nu\beta\beta$ NME of $^{48}$Ca by approximately $30\%$. Although the effects of model-space enlargement need not be the same for the two decay modes, the improved description of the observed $2\nu\beta\beta$ NME lends some support to the use of the enlarged $sdpf$ space in predicting the unobserved $0\nu\beta\beta$ NME.

	Next, we consider the $2\nu\beta^+$EC/$2\nu$ECEC candidate $^{58}$Ni in the full $fp$ shell with the GXPF1A interaction, where the participating nuclei lie in the mid-shell region and have comparatively large shell-model Hamiltonian matrix dimensions. Within 60 iterations, the $2\nu\beta^+$EC/$2\nu$ECEC NME obtained with SLSF, DLSF, and BLSF are fully converged at approximately 0.1250 (see Fig.~\ref{NME_58Ni_gxpf1a}). Of the 120 states generated with DLSF, 118 are retained after the orthogonalization and rediagonalization procedure described in Sec.~\ref{sec:LSF_based}. All 120 states generated with BLSF are retained in the NME calculation. To determine the common shift applied to the $1^+$ state energies, we used the experimental excitation energy of the $1^+_1$ state, 1.050 MeV \cite{NNDC_ENSDF}. The $m$-scheme dimension of the shell-model Hamiltonian matrix for the $1^+$ states of $^{58}$Co is $8.84\times10^{8}$, even slightly larger than that of $^{48}$Sc with the SDPF-MU interaction ($7.04\times10^{8}$); thus, we calculated the NME for $^{58}$Ni by exact diagonalization using the lowest 300 intermediate $1^+$ eigenstates. As shown in Fig. \ref{NME_58Ni_gxpf1a}(d), the resulting NME is 0.1367, which is not yet fully converged, indicating that additional intermediate $1^+$ eigenstates are required to reach the converged value of 0.1250. Experimentally, only a lower limit of $1.7\times10^{22}$ yr is known for the $2\nu\beta^+$EC half-life corresponding to the $^{58}$Ni$(0^+_{\rm g.s.})\rightarrow{}^{58}$Fe$(0^+_{\rm g.s.})$ transition \cite{Rukhadze}. For the $2\nu\beta^+$EC/$2\nu$ECEC processes, the NME is theoretically identical,
	with the only difference between the two modes arising from their respective phase-space factors. Using the experimental half-life limit \cite{Rukhadze}, $G^{2\nu}_{\beta^+\rm EC}=1.11\times10^{-24}$ yr$^{-1}$ \cite{Kotila}, and $g_{\rm A}=1.27$, we obtain an upper limit of $|M_{2\nu}|^{\rm eff}<4.51$. This experimental upper limit is considerably larger than a realistic theoretical estimate and provides only a weak constraint on the NME. The quenching factor $q=0.74$ suggested in Ref.~\cite{Iwata} for the $fp$ shell reduces the shell-model NME to $0.1250q^2=0.0685$, nearly two orders of magnitude below the experimental upper limit. A substantially more sensitive measurement of the $2\nu\beta^+\mathrm{EC}$ half-life of $^{58}$Ni is therefore needed.

	We next consider $^{76}$Ge in the full $f_{5/2}pg_{9/2}$ model space, comprising the single-particle orbitals between the magic numbers 28 and 50, using the JUN45 interaction \cite{Honma_jun45}. We perform 75 iterations with SLSF, DLSF, and BLSF, which are sufficient to obtain a fully converged $2\nu\beta\beta$ NME. Of the 150 states generated with DLSF, 142 Ritz vectors are retained after the orthogonalization and rediagonalization procedure described in Sec.~\ref{sec:LSF_based}. The SLSF, DLSF, and BLSF calculations all yield a converged $2\nu\beta\beta$ NME of 0.1656 (see Fig. \ref{NME_76Ge_jun45}). The common shift for the $1^+$ state energies of $^{76}$As was determined using the experimental excitation energy of the $1^+_1$ state, 0.044 MeV \cite{NNDC_ENSDF}. We also calculate the NME using exact diagonalization with the lowest 300 $1^+$ eigenstates of $^{76}$As, obtaining 0.1611. This indicates that additional intermediate $1^+$ eigenstates are required to reach the converged value of 0.1656. Using the recently measured $2\nu\beta\beta$ half-life of $^{76}$Ge \cite{Agostini_76Ge}, $G^{2\nu}$ from Ref. \cite{Neacsu}, and $g_{\rm A}=1.27$ in Eq.~(\ref{eq1}), the experimental NME can be estimated as $|M_{2\nu}|^{\rm eff}=0.065\pm0.001$. With the quenching factor $q=0.68$ suggested for the $f_{5/2}pg_{9/2}$ model space \cite{Suhonen_ga_eff, Kumar_gA}, the quenched NME, $0.1656q^2$, is 0.0766, which is 17.8\% larger than the experimental estimate. A previous shell-model calculation with the JUN45 interaction employed several levels of model-space truncation and obtained an unquenched $2\nu\beta\beta$ NME of 0.168 \cite{Kostensalo_76Ge}. In that calculation, the experimental energies of the three lowest intermediate $1^+$ states were used \cite{NNDC_ENSDF}, and a common shift was applied to the energies of the remaining $1^+$ states accordingly. Adopting the same intermediate-state energy prescription, the fully converged NME obtained with SLSF, DLSF, and BLSF is 0.1688, in close agreement with the value obtained in the truncated calculation \cite{Kostensalo_76Ge}.

	Finally, we consider $^{130}$Te in the full $g_{7/2}dsh_{11/2}$ model space, comprising the single-particle orbitals between the magic numbers 50 and 82, using the SN100PN interaction \cite{Brown1}. All three methods, SLSF, DLSF, and BLSF, employ 60 iterations, which are enough to achieve complete convergence of the $2\nu\beta\beta$ NME to 0.0795 (see Fig. \ref{NME_130Te_sn100pn}). Of the 120 states generated with DLSF, 114 Ritz vectors are retained after the orthogonalization and rediagonalization procedure described in Sec.~\ref{sec:LSF_based}. In all three calculations, the calculated energies of the intermediate $1^+$ states are used, as the lowest $1^+$ state in $^{130}$I has not been experimentally identified \cite{NNDC_ENSDF}. For comparison, we first evaluate the NME by exact diagonalization using the lowest 500 $1^+$ eigenstates of $^{130}$I. This calculation yields an NME of 0.0464, which is not yet converged. In this calculation, we obtain the intermediate $1^+$ eigenstates up to an excitation energy of approximately 4.3 MeV. Since calculating a few hundred additional $1^+$ eigenstates of $^{130}$I is computationally manageable, we next consider 1000 eigenstates, yielding an NME of 0.0487, as shown in Fig. \ref{NME_130Te_sn100pn}(d). Increasing the number of $1^+$ eigenstates from 500 to 1000 extends the energy range by only 0.7 MeV, up to approximately 5.0 MeV. These results indicate that a substantially larger number of intermediate $1^+$ eigenstates is required to reach the converged NME of 0.0795. This slow convergence reflects the much higher $1^+$ level density in $^{130}$I than that in the lighter odd-odd nuclei. Using the most recent experimental $2\nu\beta\beta$ half-life of $^{130}$Te from \cite{Adams}, $G^{2\nu}$ from \cite{Neacsu}, and $g_{\rm A}=1.27$, we obtain $|M_{2\nu}|^{\rm eff}=0.0187\pm0.0002$. With the quenching factor $q=0.57$ from Refs. \cite{Suhonen_ga_eff,Caurier}, suggested for the $g_{7/2}dsh_{11/2}$ model space, our calculated NME, $0.0795q^2$, becomes 0.0258. For comparison, Ref.~\cite{Horoi_LSF} reported an NME of $0.0352~{\rm MeV}^{-1}$, calculated with the GCN5082 interaction \cite{Caurier_GCN5082} in the same model space using $q=0.5$. Multiplying this value by the electron mass to adopt the dimensionless convention used here gives $M_{2\nu}=0.0180$, in close agreement with the experimental estimate. Applying the same quenching factor $q=0.5$ to our unquenched NME gives 0.0199, which is also in reasonable agreement with the experimental estimate.
	
	In addition to $^{48}$Ca, $^{58}$Ni, $^{76}$Ge, and $^{130}$Te, we tested our methods for several other candidates, including the $2\nu\beta\beta$ NMEs in $^{82}$Se with the JUN45 interaction and $^{136}$Xe with the SN100PN interaction. In all cases, the three methods yield fully converged NMEs with a relatively small number of iterations.

	\section{\label{sec:level5}Energy-denominator-free sum of the NME versus direct double-GT transition}
	
	\begin{figure*}
		\includegraphics[width=50.0mm]{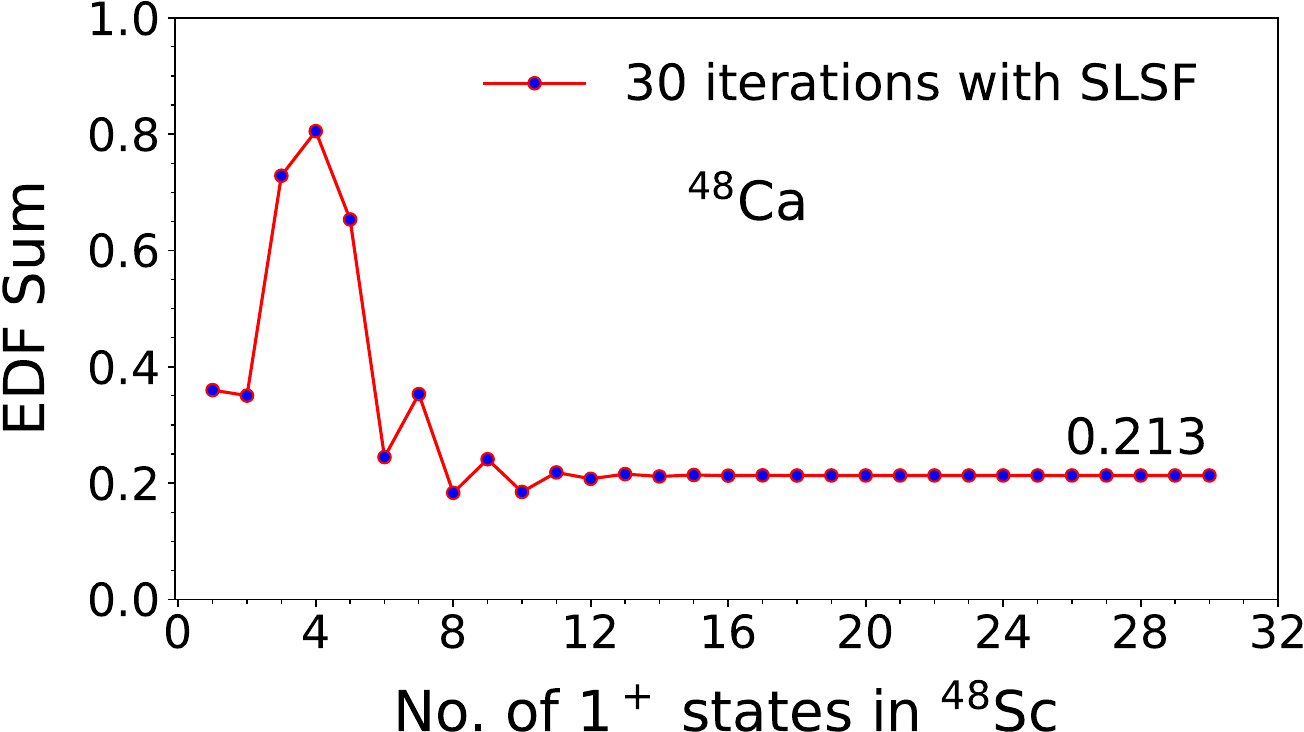}
		\includegraphics[width=46.3mm]{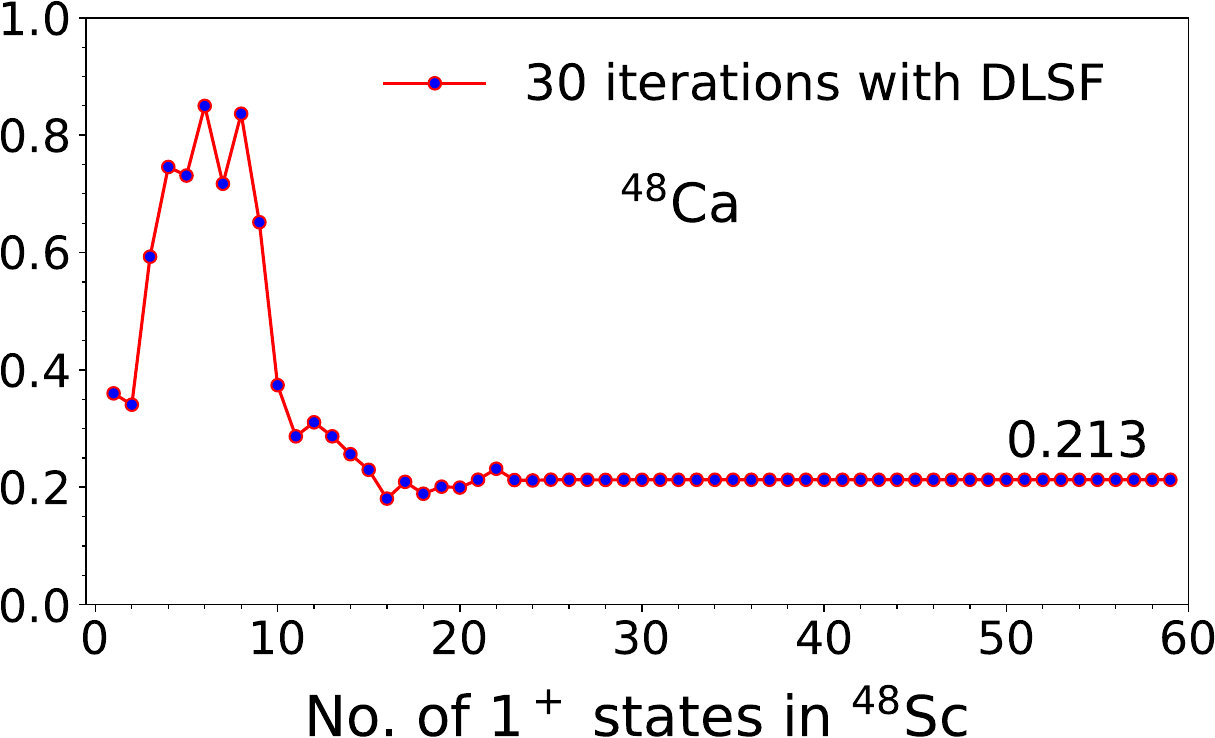}
		\includegraphics[width=46.3mm]{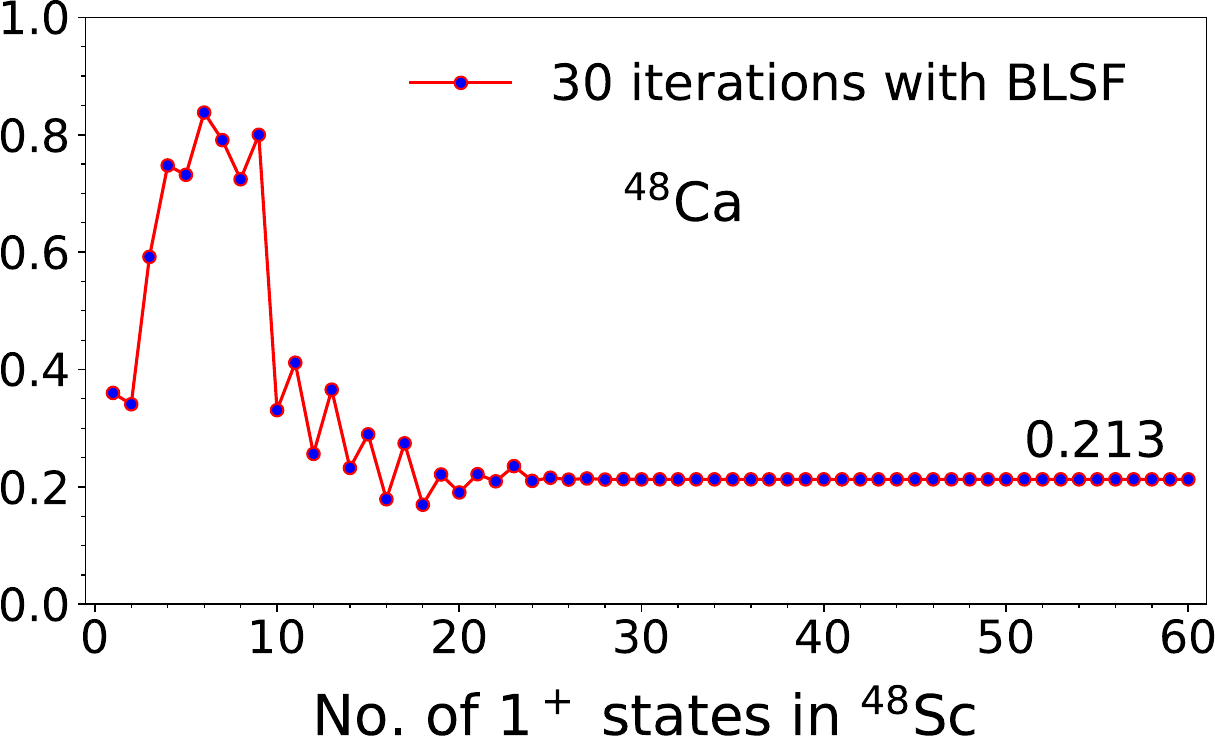}
		\includegraphics[width=50.0mm]{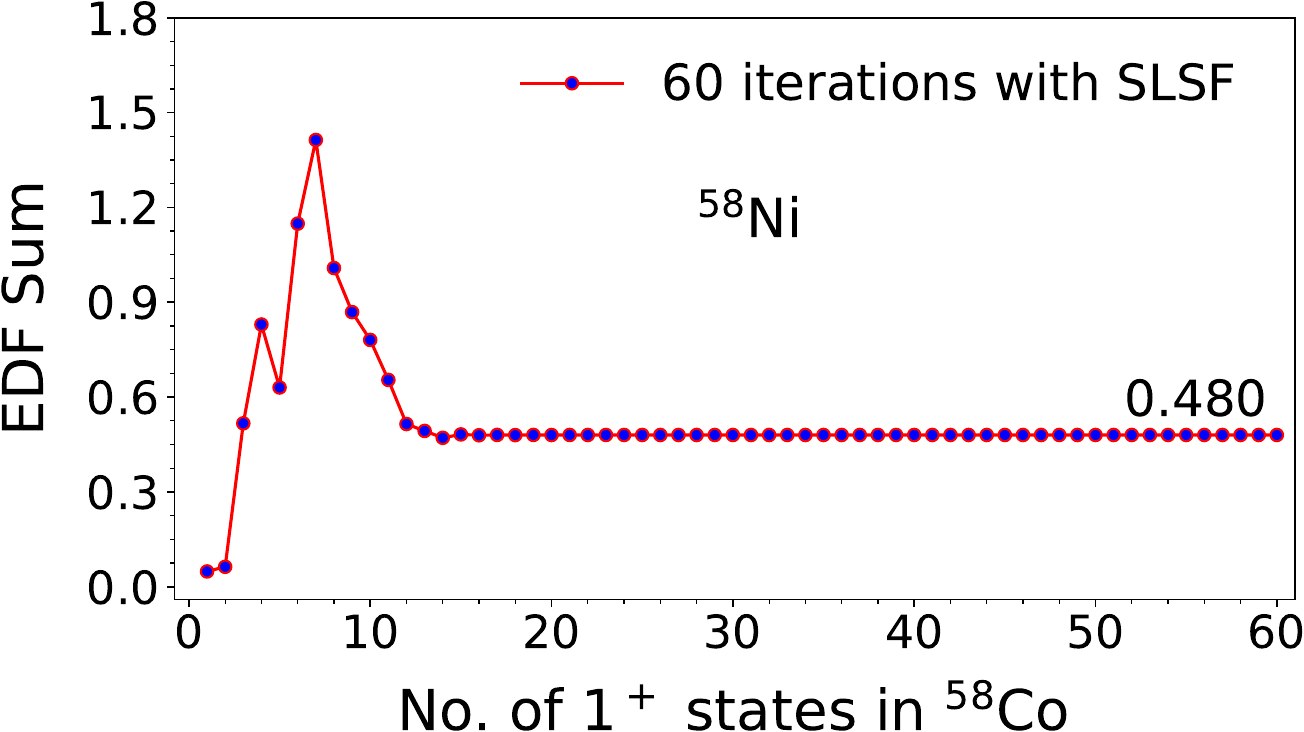}
		\includegraphics[width=46.3mm]{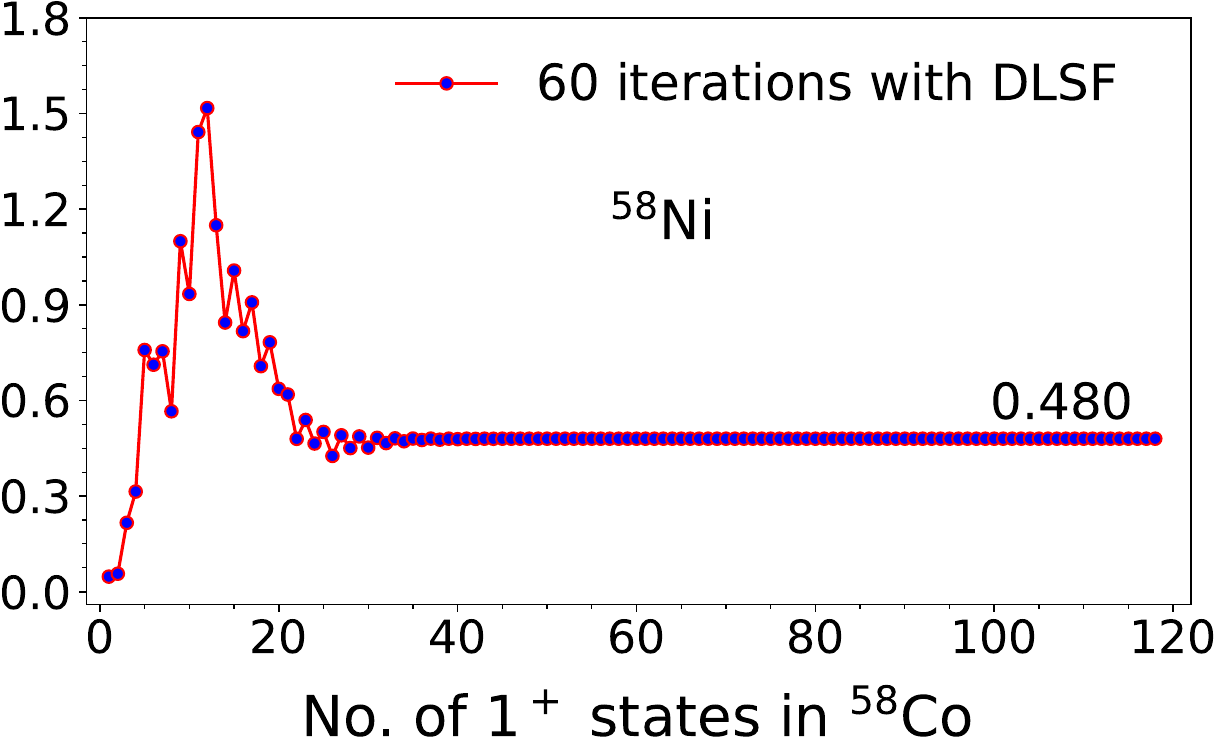}
		\includegraphics[width=46.3mm]{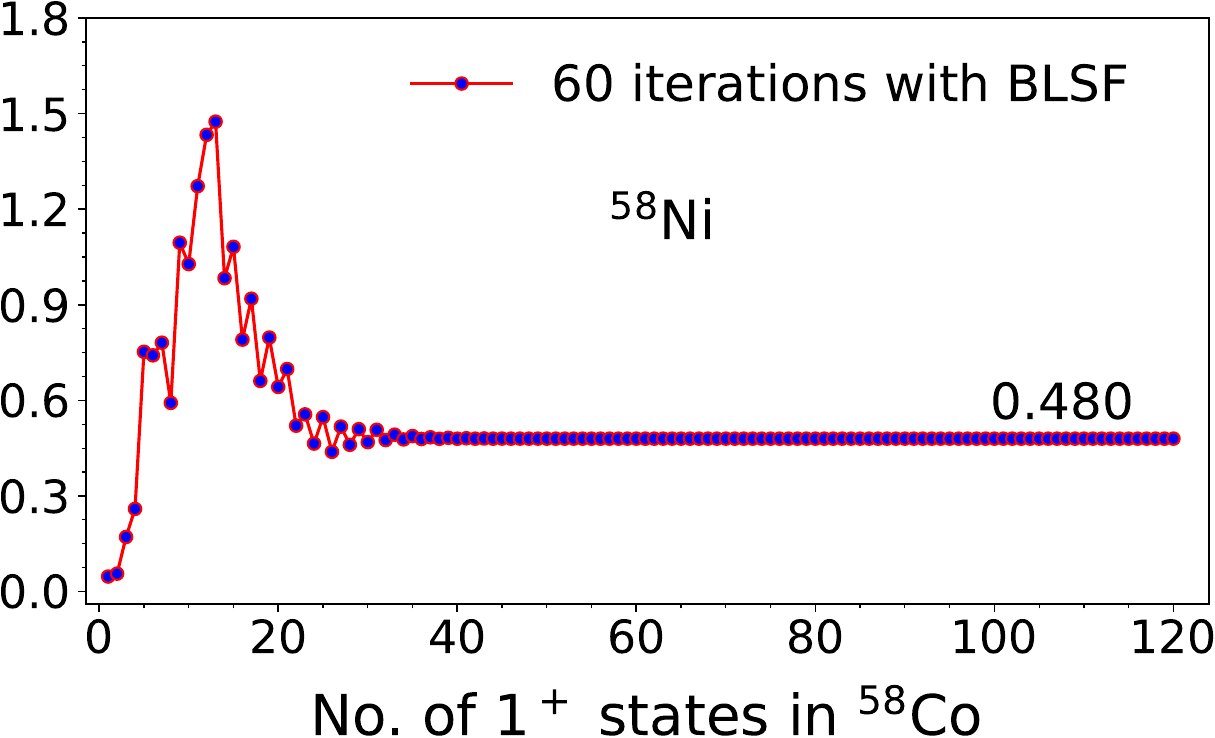}
		\includegraphics[width=50.0mm]{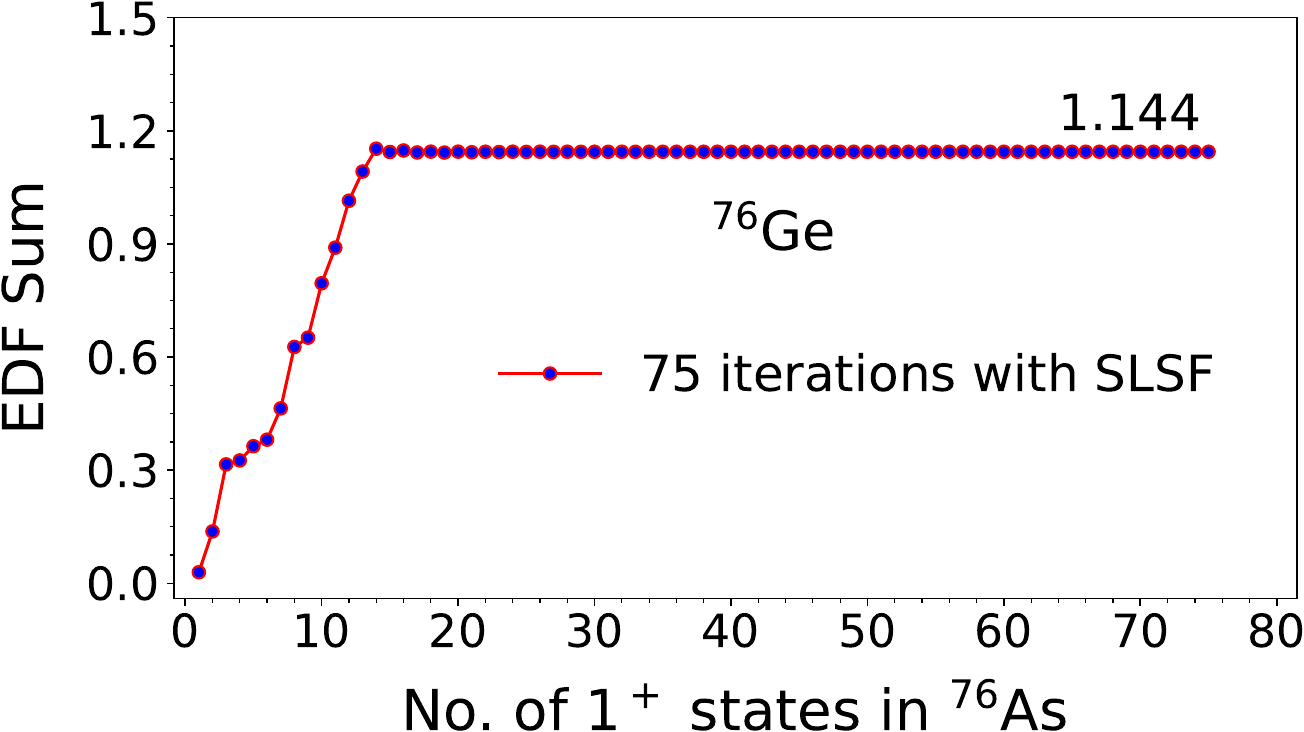}
		\includegraphics[width=46.3mm]{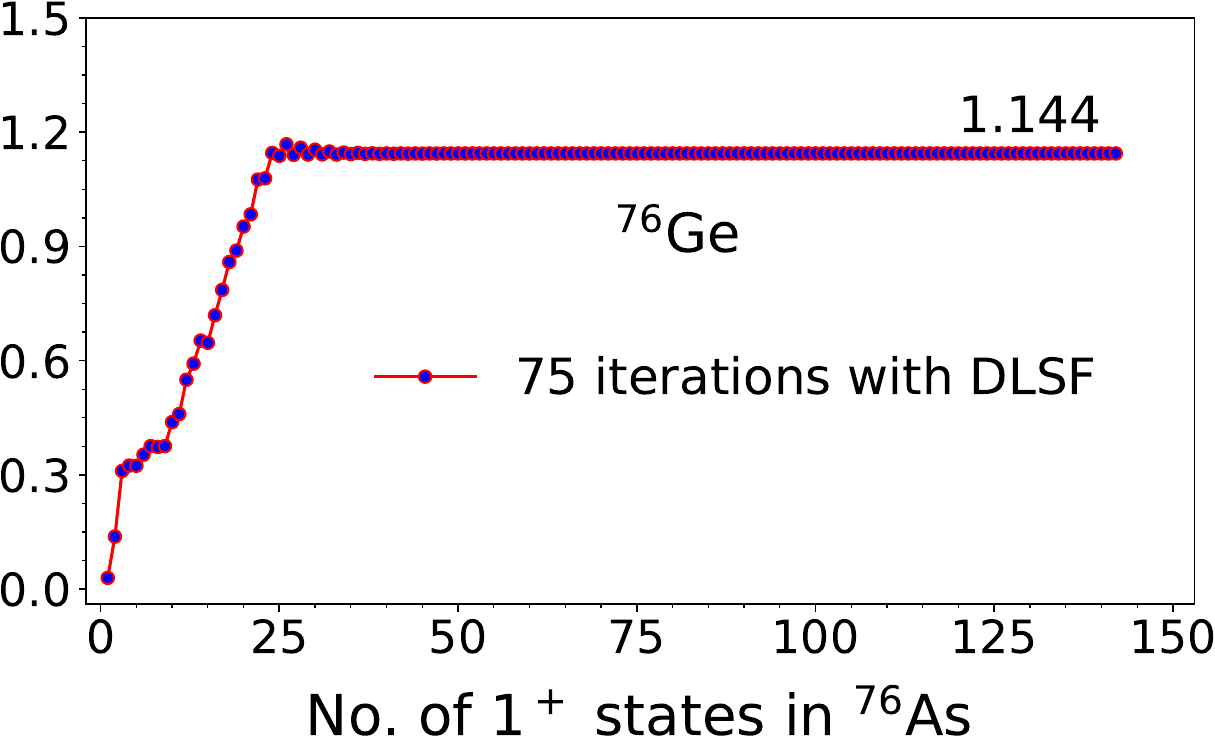}
		\includegraphics[width=46.3mm]{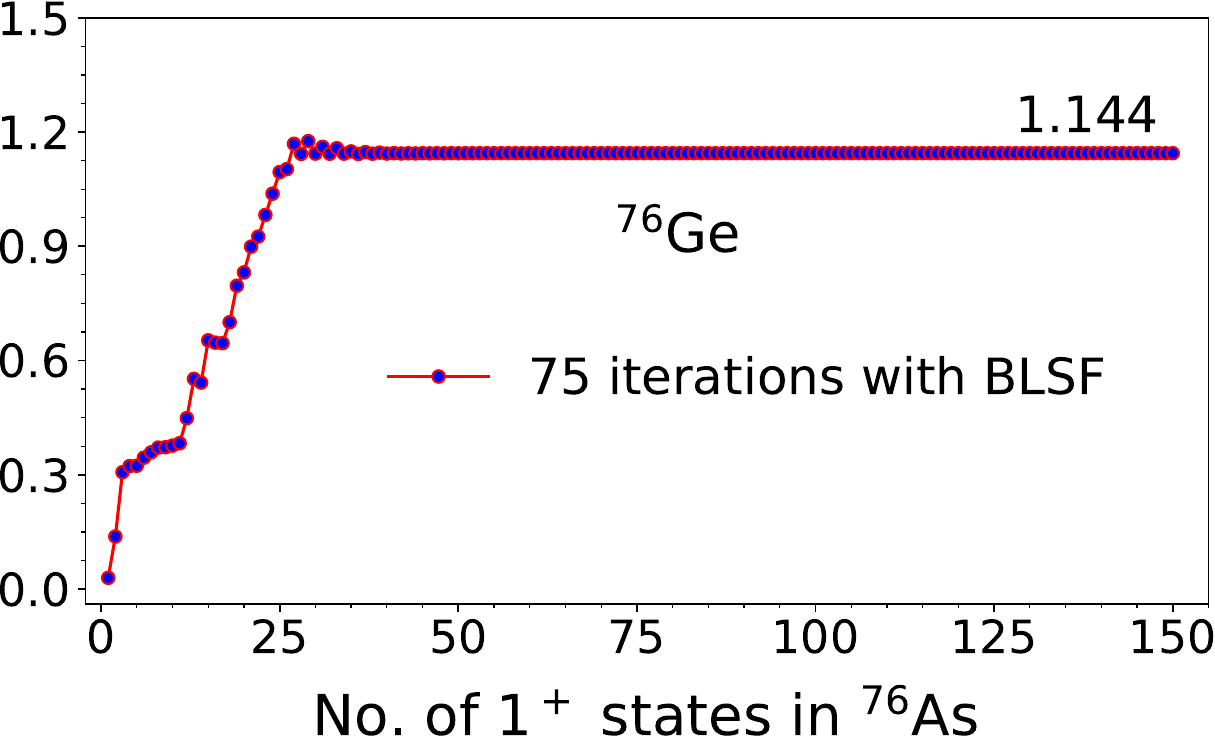}
		\includegraphics[width=50.0mm]{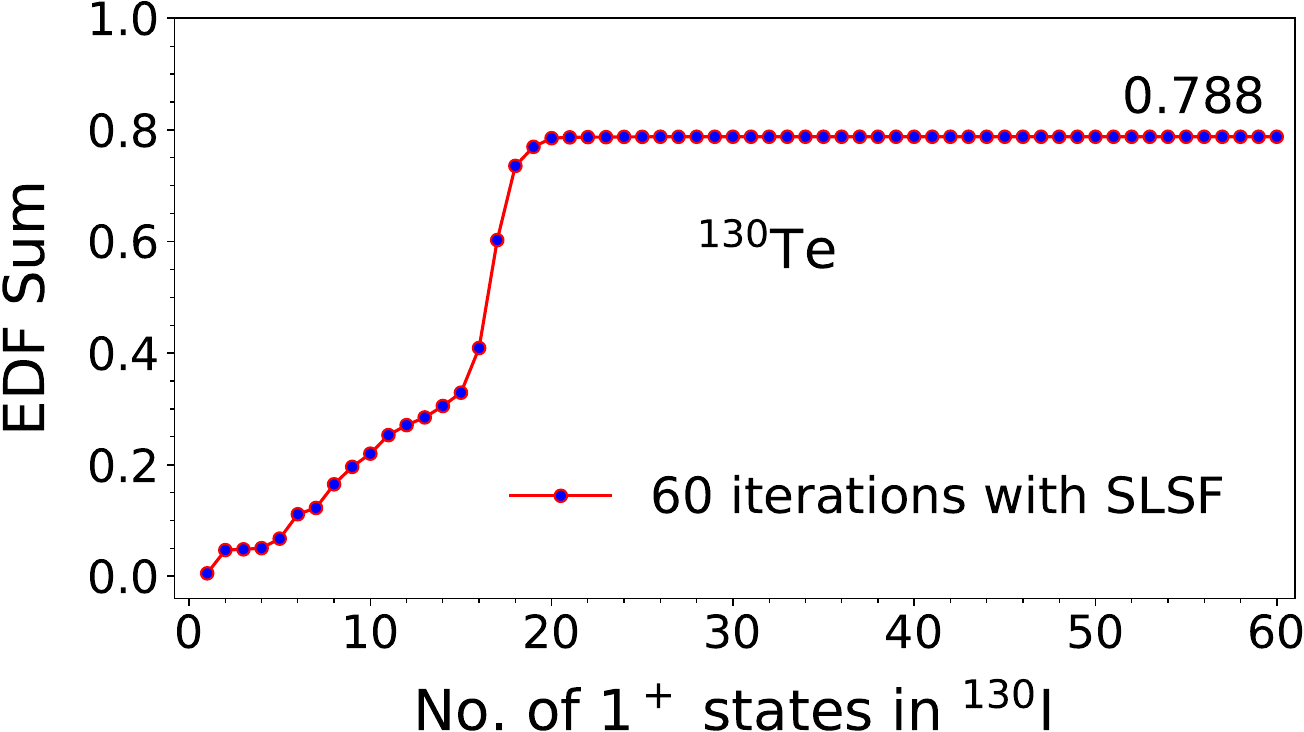}
		\includegraphics[width=46.3mm]{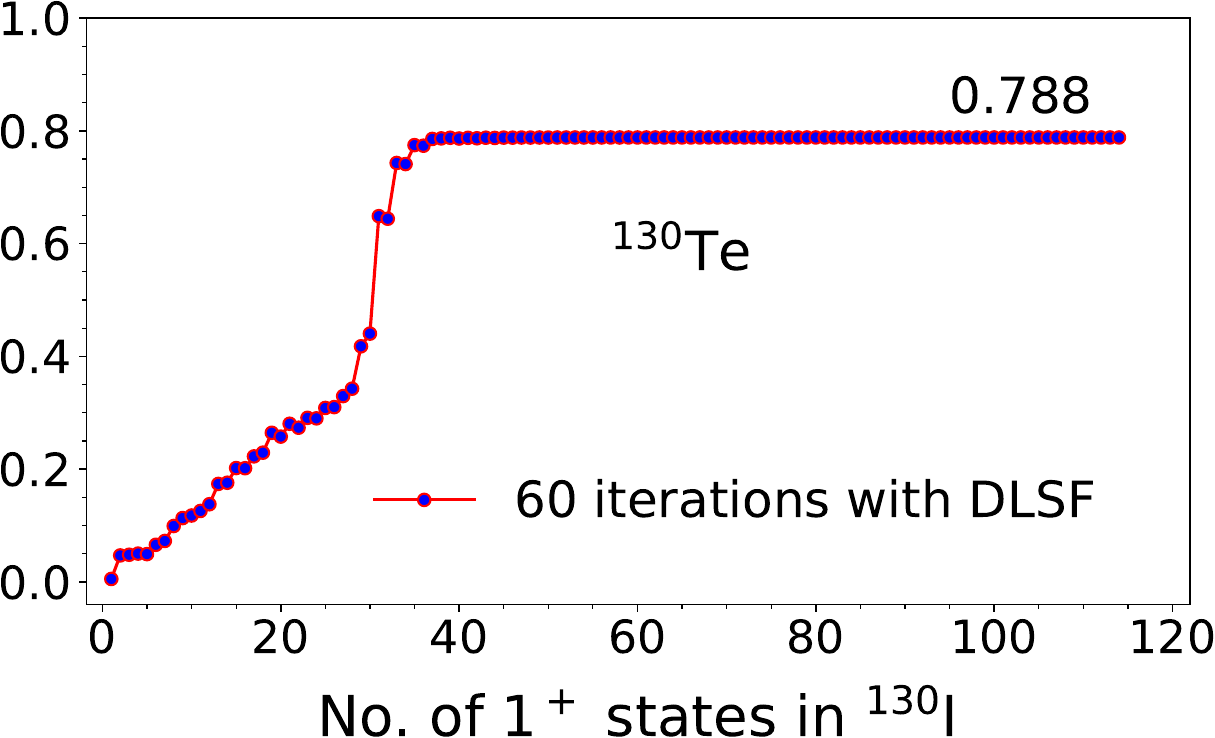}
		\includegraphics[width=46.3mm]{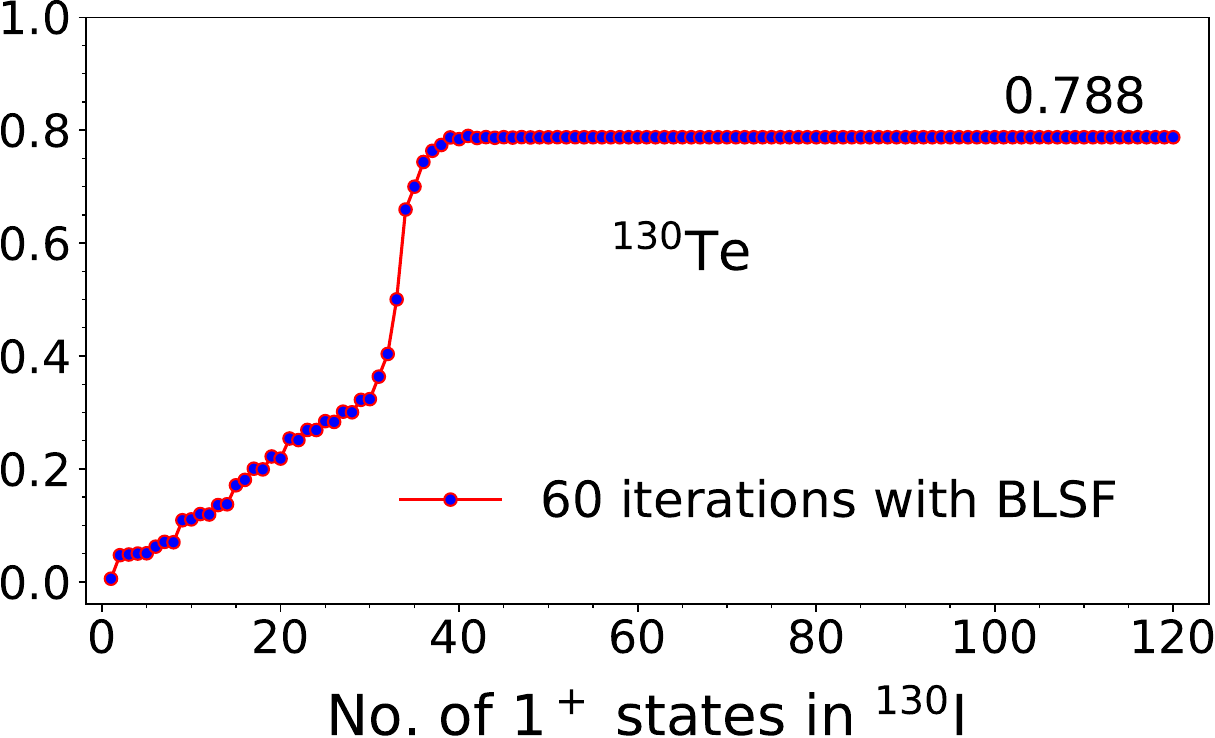}
		\caption{Energy-denominator-free (EDF) sums of the $2\nu\beta\beta/2\nu$ECEC NME for $^{48}$Ca (GXPF1A), $^{58}$Ni (GXPF1A), $^{76}$Ge (JUN45), and $^{130}$Te (SN100PN) obtained with SLSF, DLSF, and BLSF as a function of the number of intermediate $1^+$ states.}
		\label{Numer_sum_with_states}
	\end{figure*}

	The $2\nu\beta\beta/2\nu$ECEC NME defined in Eq.~(\ref{eq2}) appears to converge with the SLSF, DLSF, and BLSF methods. However, assessing the accuracy of the converged NME is difficult without a sufficiently large set of exact intermediate $1^+$ eigenstates. To test whether these methods correctly recover the complete intermediate-state contributions, we compare the energy-denominator-free sum with the exact DGT matrix element, which can be calculated directly as a two-body matrix element using the closure relation. The DGT matrix element for the $0^{+(i)}_{\rm g.s.}\rightarrow0^{+(f)}_{\rm g.s.}$ transition is defined as \cite{Shimizu_DGT}
	\begin{equation}
		M^{\rm DGT} = |\langle 0_{\rm g.s.}^{+(f)}||\sum_{a,b}[\boldsymbol{\sigma}_a\tau_a^\mp \times \boldsymbol{\sigma}_b\tau_b^\mp ]^0||0_{\rm g.s.}^{+(i)}\rangle|,
		\label{DGT_mat}
	\end{equation}
	where $[\boldsymbol{\sigma}_a\tau_a^\mp\times\boldsymbol{\sigma}_b\tau_b^\mp]^0$ denotes the rank-0 DGT operator. Equivalently, by inserting a complete set of intermediate $1^+$ states, this matrix element can be expressed as
	\begin{equation}
		M^{\rm DGT} = \frac{1}{\sqrt{3}} \left| \sum_n \langle 0_{\rm g.s.}^{+(f)}\|
		\hat{O}_{\rm GT}^{a} \|1_n^+\rangle \langle 1_n^+\| \hat{O}_{\rm GT}^{b}
		\|0_{\rm g.s.}^{+(i)}\rangle \right|,
		\label{DGT_sum}
	\end{equation}
	where $\hat{O}_{\rm GT}^{a}=\sum_a\boldsymbol{\sigma}_a\tau_a^\mp$ and $\hat{O}_{\rm GT}^{b}=\sum_b\boldsymbol{\sigma}_b\tau_b^\mp$, and the sum runs over the complete set of intermediate $1^+$ states. The factor $1/\sqrt{3}$ follows from the standard angular-momentum coupling convention for the rank-0 product of two rank-1 operators \cite{Saliba}. The energy-denominator-free sum in the $2\nu\beta\beta/2\nu$ECEC NME (Eq.~(\ref{eq2})) consists of the same products of reduced GT matrix elements as in Eq.~(\ref{DGT_sum}), but without the factor $1/\sqrt{3}$. Thus, when contributions from all intermediate $1^+$ states are included, its magnitude is related to $M^{\rm DGT}$ by this normalization factor, and we can write the energy-denominator-free sum as $\sqrt{3}M^{\rm DGT}$. It should be noted that the energy denominator suppresses the contributions of highly excited intermediate $1^+$ states to the full $2\nu\beta\beta/2\nu$ECEC NME, whereas this is not the case for the energy-denominator-free sum. Comparing the cumulative energy-denominator-free sums with the direct DGT matrix element therefore provides a stringent test of whether SLSF, DLSF, and BLSF recover the complete intermediate-state contributions, independently of energy-denominator effects. Moreover, previous shell-model calculations have found a correlation between the $0^{+(i)}_{\rm g.s.}\rightarrow0^{+(f)}_{\rm g.s.}$ DGT matrix element and the $0\nu\beta\beta$ NME \cite{Shimizu_DGT}.

	Figure~\ref{Numer_sum_with_states} shows the cumulative energy-denominator-free sums obtained with SLSF, DLSF, and BLSF for $^{48}$Ca, $^{58}$Ni, $^{76}$Ge, and $^{130}$Te as functions of the number of intermediate $1^+$ states. For each nucleus, we use the same number of iterations as in the corresponding $2\nu\beta\beta/2\nu$ECEC NME calculation discussed in Secs.~\ref{sec:level3} and \ref{sec:level4}, namely, 30, 60, 75, and 60 iterations, respectively. For the $^{48}$Ca$(0^+_{\rm g.s.})\rightarrow{}^{48}$Ti$(0^+_{\rm g.s.})$ transition in the $fp$ shell with the GXPF1A interaction, the directly calculated DGT matrix element from Eq.~(\ref{DGT_mat}) gives $\sqrt{3}M^{\rm DGT}=0.213$. The cumulative energy-denominator-free sums obtained with SLSF, DLSF, and BLSF all reproduce this value, as shown in Fig.~\ref{Numer_sum_with_states}. For the $^{58}$Ni$(0^+_{\rm g.s.})\rightarrow{}^{58}$Fe$(0^+_{\rm g.s.})$, $^{76}$Ge$(0^+_{\rm g.s.})\rightarrow{}^{76}$Se$(0^+_{\rm g.s.})$, and $^{130}$Te$(0^+_{\rm g.s.})\rightarrow{}^{130}$Xe$(0^+_{\rm g.s.})$ transitions, the directly calculated DGT matrix elements give $\sqrt{3}M^{\rm DGT}=0.480$, 1.144, and 0.788, respectively. In each case, the cumulative energy-denominator-free sums obtained with all three methods reproduce these values.
	
	We have performed the same comparison for several other candidates across different mass regions and obtained consistent results. The agreement between the energy-denominator-free sums and the directly calculated DGT matrix elements demonstrates that SLSF, DLSF, and BLSF correctly recover the complete intermediate-state contributions for the cases considered here. This agreement provides an independent validation of the intermediate-state representations used in the $2\nu\beta\beta/2\nu$ECEC NME calculations.

	\section{\label{sec:level6}Summary and Conclusions}
	
	We have developed two approaches for calculating $2\nu\beta\beta/2\nu$ECEC NMEs within the nuclear shell model framework, based on the Lanczos strength-function and block-Lanczos methods. Both approaches avoid the explicit calculation of a complete set of exact intermediate $1^+$ eigenstates and yield fully converged $2\nu\beta\beta/2\nu$ECEC NMEs using a relatively small number of iterations starting from GT doorway states. Unlike the previous single-sided Lanczos strength-function approach, the present DLSF and BLSF methods simultaneously preserve the total GT strengths associated with both the initial and final nuclei. The methods were first benchmarked for $^{48}$Ca in the $fp$ shell using the GXPF1A interaction and were subsequently applied to $2\nu\beta\beta/2\nu$ECEC candidates in different mass regions.
	
	Specifically, the methods were applied to $^{48}$Ca, $^{58}$Ni, $^{76}$Ge, and $^{130}$Te in the $sdpf$, $fp$, $f_{5/2}pg_{9/2}$, and $g_{7/2}dsh_{11/2}$ model spaces using the SDPF-MU, GXPF1A, JUN45, and SN100PN interactions, respectively. For $^{48}$Ca, including cross-shell excitations of up to $2\hbar\omega$ in the $sdpf$ model space reduces the calculated $2\nu\beta\beta$ NME relative to the full-$fp$-shell result and brings it closer to the experimental estimate. This result indicates that enlarging the model space improves at least some aspects of the nuclear-structure description relevant to $\beta\beta$ decay in $^{48}$Ca. For $^{58}$Ni, we have presented, to our knowledge, the first shell-model calculation of the $2\nu\beta^+$EC/$2\nu$ECEC NME. The theoretical $2\nu\beta^+$EC NME is approximately two orders of magnitude smaller than the current experimental upper limit, indicating the need for a more sensitive measurement of the $2\nu\beta^+$EC half-life.
	
	As an independent validation of the intermediate-state representations, we compared the energy-denominator-free sums obtained with SLSF, DLSF, and BLSF with the corresponding DGT matrix elements calculated directly between the initial and final ground states. For all nuclei considered, the three methods reproduce the corresponding direct DGT matrix elements. This agreement demonstrates that the intermediate-state representations generated by these methods recover the complete energy-denominator-free sums under the same calculation conditions used for the $2\nu\beta\beta/2\nu$ECEC NMEs.
	
	The Lanczos strength-function method has previously been applied to evaluate the ultrasoft-neutrino contribution to the $0\nu\beta\beta$ amplitude of $^{136}$Xe \cite{Dekens2023}. This previous application provides a basis for exploring extensions of the DLSF and BLSF approaches to $0\nu\beta\beta$ decay. Calculation of the full $0\nu\beta\beta$ NME would, however, require an appropriate treatment of the momentum-dependent neutrino potential in the transition operator.

	\begin{acknowledgments} 
		
		We acknowledge support from JSPS KAKENHI (Grant No. 26K24536) and thank the D3 Center at Osaka University for providing computing resources on SQUID and the Miyabi supercomputer system at the Information Technology Center, The University of Tokyo. We also acknowledge support from JSPS KAKENHI (Grant Numbers 25K07330 and 25K00995) and the MCRP program (NUCLSM) at the Center for Computational Sciences, University of Tsukuba.
		
	\end{acknowledgments}
	
	


\end{document}